\documentclass[12pt,a4paper,amsmath,amssymb]{article}
\usepackage[margin=0.8in]{geometry}
\usepackage{longtable}
\usepackage{amssymb}
\usepackage{amsmath}
\usepackage{amsfonts}
\usepackage{bm}
\usepackage[utf8]{inputenc}
\usepackage{graphicx}
\usepackage{float}
\usepackage{xcolor}
\usepackage{caption}
\usepackage{multirow}
\definecolor{bluemunsell}{rgb}{0.0, 0.5, 0.69}
\usepackage[colorlinks,linkcolor = bluemunsell,
urlcolor  = bluemunsell,
citecolor = bluemunsell,
anchorcolor = bluemunsell]{hyperref}
\usepackage{MnSymbol,bbding,pifont}
\usepackage{xcolor}
\usepackage{makecell}
\usepackage{ctable} 
\usepackage{adjustbox}
\def\thefootnote{\fnsymbol{footnote}}

\usepackage{cite}

\begin{document}
	\begin{center}
		{\Large \textbf{
				Searching for $H \to hh \to b\bar b\tau\tau$ \\[0.25cm] in the 2HDM Type-I at the LHC
		}} \\
		\thispagestyle{empty} 
		\vspace{1cm}
		{\sc
			A. Arhrib$^{1,2}$\footnote{\href{mailto:a.arhrib@gmail.com}{a.arhrib@gmail.com}},
			S. Moretti$^{3,4}$\footnote{\href{mailto:s.moretti@soton.ac.uk}{s.moretti@soton.ac.uk}; \href{mailto:stefano.moretti@physics.uu.se}{stefano.moretti@physics.uu.se}},
			S. Semlali$^{3,5}$\footnote{\href{mailto:souad.semlali@soton.ac.uk}{souad.semlali@soton.ac.uk}},
			C.~H.~Shepherd-Themistocleous$^{5}$\footnote{\href{mailto:claire.shepherd@stfc.ac.uk}{claire.shepherd@stfc.ac.uk}},
			Y. Wang$^{6,7}$\footnote{\href{mailto:wangyan@imnu.edu.cn}{wangyan@imnu.edu.cn}},
			Q.S. Yan$^{8,9}$\footnote{\href{mailto:yanqishu@ucas.ac.cn}{yanqishu@ucas.ac.cn}}\\
		}
		\vspace{1cm}
		{\sl
			$^1$ {Abdelmalek Essaadi University, Faculty of Sciences and Techniques, B.P. 2117 T\'etouan, Tanger, Morocco.}\\
			\vspace{0.2cm}
			$^2$ {Department of Physics and Center for Theory and Computation, National Tsing Hua University, Hsinchu, Taiwan 300.}\\
			\vspace{0.2cm}
			$^3$\small{School of Physics and Astronomy, University of Southampton, Southampton, SO17 1BJ,\\ United Kingdom.}\\
			\vspace{0.2cm}
			$^4$\small{Department of Physics and Astronomy, Uppsala University, Box 516, SE-751 20 Uppsala, Sweden.}\\
			\vspace{0.2cm}	
			$^5$\small{Particle Physics Department, Rutherford Appleton Laboratory, Chilton, Didcot, Oxon OX11 0QX, United Kingdom.}\\
			\vspace{0.2cm}
			$^6$\small{College of Physics and Electronic Information, Inner Mongolia Normal University,\\ Hohhot 010022, PR China.}\\
			\vspace{0.2cm}
			$^7$\small{Inner Mongolia Key Laboratory for Physics and Chemistry of Functional Materials,\\
				Inner Mongolia Normal University, Hohhot, 010022, China.}\\
			\vspace{0.2cm}
			$^8$\small{Center for Future High Energy Physics, Chinese Academy of Sciences,\\ Beijing 100049, P.R. China.}\\
			\vspace{0.2cm}
			$^9$\small{School of Physics Sciences, University of Chinese Academy of Sciences,\\ Beijing 100039, P.R. China.
		}}
	\end{center}
	
	\begin{abstract}
		
		Unlike other realisations of the 2-Higgs Doublet Model (2HDM), the so-called Type-I allows for a very light Higgs boson spectrum. Specifically, herein, the heaviest of the two CP-even neutral Higgs states, $H$, can be the one discovered at the Large Hadron Collider (LHC) in 2012, with a mass of $\approx 125$ GeV and couplings consistent with those predicted by the Standard Model (SM). In such a condition of the model, referred to as `inverted mass hierarchy', the decay of the  SM-like Higgs state into pairs of the lightest CP-even neutral Higgs boson, $h$, is possible, for masses of the  latter ranging from $M_H/2\approx 65$ GeV down to 15 GeV or so, all compatible with experimental constraints. In this paper, we investigate the scope of the LHC in accessing the process $gg\to H \to hh\to b\bar b\tau\tau$ by performing a Monte Carlo (MC) analysis aimed at extracting this signal from the SM backgrounds, in presence of a dedicated trigger choice and kinematic selection. We prove that some sensitivity to such a channel exists already at Run 3 of the LHC while the High-Luminosity LHC (HL-LHC) will be able to either confirm or disprove this theoretical scenario over sizable regions of its parameter space.  	
	\end{abstract}

	\def\thefootnote{\arabic{footnote}}
	\setcounter{page}{0}
	\setcounter{footnote}{0}
	
	\newpage
	\section{Introduction}

	In the SM of particle physics, it is well known that the Higgs boson  \cite{ATLAS:2012yve,CMS:2012qbp} is responsible 
	for the generation of fermion and gauge boson masses through  what is called Spontaneous Symmetry Breaking (SSB) \cite{Englert:1964et,Higgs:1964pj}. Such a mechanism also predicts a self-interaction for the Higgs state.
 
	The measurement of such a self-coupling is the only experimental way to understand the SSB mechanism and to reconstruct the Higgs potential responsible for it.  This is an important (and challenging) task also because it can shed some light on possible  Beyond the SM (BSM)  effects that may affect Higgs self-couplings in general.
	
	The LHC has started a new campaign of measurements after the recent upgrade, the so-called Run 3. This will involve, among other things, measuring ever more precisely the coupling of the SM-like Higgs boson to other SM particles or even progressing towards the measurement of its self-coupling. The LHC is also capable of measuring new decays of the SM-like Higgs boson into non-SM particles.
	Current results from the ATLAS and CMS experiments indicate 
	that the measured SM-like Higgs signal rates in all  channels agree well  with the SM theoretical predictions at the
	$\sim 2\sigma$ level~\cite{ATLAS:2022vkf,CMS:2022dwd}. 
	However, there are several pieces of evidences, both theoretical (the hierarchy problem, the absence of gauge coupling unification, etc.) and experimental (neutrino masses, the matter-antimatter asymmetry, etc.), which indicate that the SM could not be the 
	ultimate description of Nature  but should be viewed as a low-energy effective theory of some more  fundamental  one yet to be discovered.
	
	There exist several BSM theories that address these weaknesses of the SM while identifying the 125 GeV scalar particle as a part of an extended scalar sector.
	One of the simplest extensions of the SM is the 2HDM, which contains
	two Higgs doublets, $\Phi_1$ and $\Phi_2$, which give masses to all 
	fermions and gauge bosons. The particle spectrum of the 2HDM
	is as follows: two CP-even ($h$ and $H$, with $m_h<m_H$, one of them being identified with the 
	SM-like Higgs boson with mass 125 GeV: $H$ in our case),  one CP-odd ($A$) and a pair of charged  ($H^\pm$) Higgs bosons. 
	
	According to the latest experimental results  from both ATLAS and CMS,  the presence of non-SM decay modes of the SM-like
	Higgs boson is not completely ruled out. Both experiments have set upper limits on the Branching Ratio (BR) of such non-SM decays which are 12\% for ATLAS \cite{ATLAS:2022vkf} and 16\% for CMS \cite{CMS:2022dwd}.
	The LHC experiments are expected to soon constrain the BRs of such non-SM decays beyond the 5-10\% level using indirect measurements \cite{Liss:2013hbb,CMS:2013xfa}. 
	There exist several BSM models that possess such non-SM decays of the SM-like Higgs boson: non-minimal scenarios of Supersymmetry \cite{Moretti:2019ulc} such as 
	the Next-to-Minimal Supersymmetric Standard Model and new Monimal Supersymmetric Standard Model (NMSSM/mMSSM) \cite{Dedes:2000jp,Dobrescu:2000yn,Ellwanger:2003jt,Dermisek:2005ar}, models for Dark Matter (DM)  \cite{Pospelov:2007mp,Draper:2010ew,Ipek:2014gua,Arhrib:2013ela}, scenarios with first order Electro-Weak (EW) phase transitions \cite{Profumo:2007wc,Blinov:2015sna} and an  extended Higgs sector \cite{Lee:1973iz,Branco:2011iw}.
	It is then crucial to use LHC Higgs measurements to test  BSM models 
	that predict such exotic SM-like Higgs decays (i.e., into non-SM particles). 
	In the 2HDM, if the heavy CP-even $H$ is the observed SM-like Higgs boson, then $H$ can decay into a pair of light 
	CP-even Higgs states, $H\to hh$, or CP-odd ones, $H\to AA$. The phenomenology of such decays of the observed SM-like Higgs boson is studied in Refs.~\cite{Arhrib:2017uon,Celis:2013rcs,Grinstein:2013npa,Chen:2013rba} for the case of the 2HDM, with an emphasis on the so-called Type-I (see below).
	
	Following the discovery of the SM-like Higgs boson by ATLAS and CMS at the LHC in 2012, there have been several experimental searches for exotic decays of the SM-like Higgs through four fermions final states: $pp\to H \to XX\to f_1 f_1 f_2f_2$, where $f_{1,2}$ are 2 light fermions such as muons, taus or bottom quarks. Such a search clearly benefits  from the large cross-section $\sigma(pp\to H)$ as well as from the large BR of the exotic SM-like Higgs decay $H\to hh$ that could reach up to 10\% in some BSM scenarios. In that spirit, the ATLAS and CMS collaborations have performed several searches looking for 
	$\tau\tau\tau\tau$ \cite{CMS:2017dmg,CMS:2019spf}, $\tau\tau$$\mu\mu$ \cite{ATLAS:2015unc,CMS:2017dmg,CMS:2018qvj,CMS:2020ffa}, $b\overline{b}\mu\mu$ \cite{CMS:2017dmg,CMS:2022fxg,CMS:2023ryd}, $b\overline{b}\tau\tau$ \cite{CMS:2017hea,CMS:2018zvv}, $eeee$ \cite{ATLAS:2018coo}, $ee\mu\mu$ \cite{ATLAS:2018coo} and $\mu\mu\mu\mu$ \cite{CMS:2012qms,CMS:2015nay,ATLAS:2018coo,CMS:2018jid,ATLAS:2021ldb}, 
	which have enabled one to set an upper limit on the BR of the $X$ decay into any given 2 light fermions.

	Motivated by the recent search for $b\overline{b}\tau\tau$ final states conducted by the CMS experiment \cite{CMS:2017hea,CMS:2018zvv}, herein, we would like 
	to address the study of signal and background for such a final state within the so-called 2HDM Type-I. We first demonstrate that, 
	within this framework, the production rate of $b\overline{b}\tau\tau$ via the process $\sigma(pp\to H) \times {\rm BR}(H\to hh) \times {\rm BR}(h\to b\overline{b})\times {\rm BR}(h\to \tau \tau)$ could be substantial and then perform a feasibility study based on a signal-versus-background analysis using standard Monte Carlo (MC) simulation tools. It is found that the interesting parameter space could be either confirmed or disproved by a large data set (say 3000 fb$^{-1}$), which can be attained at the High Luminosity LHC (HL-LHC) \cite{Gianotti:2002xx}, with possible hints of such a signal already at Run 3 of the LHC (for 300 fb$^{-1}$).
	
	The paper is organised as follows: in section 2 we give a brief review of the 2HDM 
	and describe the theoretical and experimental constraints that are used. In section 3, we present some general features of the SM-like Higgs boson decays into two light Higgs bosons and further into $b\overline{b}\tau\tau$ final states over the parameter space of the inverted mass hierarchy scenario of the 2HDM Type-I. In section 4, we perform a detailed MC study of the feasibility of the signal process $ g g \to H \to h h \to b\bar{b} \tau \tau$ at the current Run 3 of the LHC and future HL-LHC  stages. In section 5, we end this work with some conclusions.
	
	\section{Theoretical and Experimental Constraints}
	\label{section1}
	The 2HDM is obtained by extending the Higgs sector of the SM with an additional Higgs doublet field. 
	Assuming that $\Phi_i$ (${i=1,2}$)  are the two $SU(2)_L$  Higgs doublets and $v_{1,2}$ are their Vacuum Expectation Values (VEVs),  the most general renormalisable potential which is invariant under 
	${SU(2)_L \times U(1)_Y}$ is given by \cite{Branco:2011iw}:
	\begin{eqnarray}
	V_{\rm{Higgs}}(\Phi_1,\Phi_2) &=& \lambda_1(\Phi_1^\dagger\Phi_1)^2 +
	\lambda_2(\Phi_2^\dagger\Phi_2)^2 +
	\lambda_3(\Phi_1^\dagger\Phi_1)(\Phi_2^\dagger\Phi_2) +
	\lambda_4(\Phi_1^\dagger\Phi_2)(\Phi_2^\dagger\Phi_1)  +\\ && +
	\frac12\left[\lambda_5(\Phi_1^\dagger\Phi_2)^2 +\rm{h.c.}\right] 
	+m_{11}^2 \Phi_1^\dagger \Phi_1+ m_{22}^2\Phi_2^\dagger
	\Phi_2 + \left[m_{12}^2
	\Phi_1^\dagger \Phi_2 - \rm{h.c.}\right] \,. \nonumber \label{2hdmpot}
	\end{eqnarray}
	By hermiticity, $\lambda_{1,2,3,4}$ as well as $m_{11}^2$ and $m_{22}^2$ are all real-valued.
	The parameters $\lambda_{5}$ and $m_{12}^2$ can be complex and can generate CP violation in the Higgs sector. However, in what follows, we assume that there is no such phenomenon.
	
	After EW Symmetry Breaking (EWSB) takes place, from the 8 degrees of freedom initially present in $\Phi_1$ and $\Phi_2$, 3 are taken up by the ensuing  Goldstone bosons to give masses to the gauge bosons $W^\pm$ and $Z$, so 
	we are eventually left with 5  physical Higgs states. A pair of charged Higgs $H^\pm$, a CP-odd $A$ and two CP-even states $H$ and $h$  (with $m_{h}< m_{H}$), as mentioned.
	One of the neutral CP-even Higgs states has to  be identified with the 125 GeV SM-like Higgs particle observed at the LHC. In the present study, as intimated, we will assume $m_H=125$ GeV while $m_h<m_H/2$.
	
	The whole Higgs sector of the 2HDM is then parameterised  by 7 parameters: e.g., 
	\begin{eqnarray}
	m_{H^{\pm}},~m_{A},~m_{H},~m_{h},~\alpha,~\beta ~\text{and} \ m_{12}^2,
	\label{eq:param} 
	\end{eqnarray}
	where $\alpha$ is the mixing angle between the CP-even components of the neutral Higgs states of the doublet fields while
	$\beta$ is the mixing between the CP-odd components and is given by $\tan\beta=v_2/v_1$. 
	As we are considering the scenario where $H$ state is the 125 GeV scalar, the SM (alignment) limit is recovered when $\cos(\beta-\alpha) \approx 1$.
	
	From the above Higgs potential, one can derive, in particular, the triple Higgs coupling $Hhh$ needed for our analysis, as follows  \cite{Arhrib:2008jp,Kanemura:2015mxa}:
	\begin{eqnarray}
	Hhh&=& -\frac{g c_{\beta-\alpha}}{2m_Ws_{2\beta}^2} 
	\left[(2m_h^2+m_H^2) s_{2\alpha} s_{2\beta} - 2( 3 s_{2\alpha} - s_{2 \beta}) m_{12}^2 \right], 
	\label{tri-coupling}
	\end{eqnarray}
	where $s_x$ and $c_x$ are shorthand notations for $\sin x$ and $\cos x$, respectively. The coupling $Hhh$ is  proportional to $\cos(\beta-\alpha)$ which is close to unity in our scenario with $H$ being the observed SM-like Higgs boson, as explained.

	In the Yukawa sector, if we proceed to EWSB like in the SM, we end up with Flavour Changing Neutral Currents (FCNCs)  at tree level.
	Such dangerous FCNCs can, however, be avoided by imposing a discrete $Z_2$ symmetry  \cite{Glashow:1976nt}
	by coupling each fermion type to only one of the Higgs doublets. As a consequence, 
	there are four types of 2HDM \cite{Branco:2011iw}, of which,  in this study, we are interested only in the so-called Type-I, wherein only the doublet 
	$\Phi_2$ couples to all fermions exactly as in the SM \cite{Branco:2011iw}.

	The neutral Higgs couplings  to fermions can be obtained from the Yukawa Lagrangian and are given by \cite{Branco:2011iw}: 
	\begin{eqnarray}
	- {\mathcal{L}}_{\rm Yukawa} &=& \sum_{f=u,d,l} \frac{m_f}{v}  \left(   \frac{\cos\alpha}{\sin\beta} \bar{f} f h+ 
	\frac{\sin\alpha}{\sin\beta}  \bar{f} f H\right).
	\label{Yukawa-1}
	\end{eqnarray}
	As one can read from the above Lagrangian term, the CP-even neutral Higgs couplings to quarks and leptons are similar in the 2HDM Type-I, since both  are proportional to $\frac{1}{\sin\beta} \propto \frac{1}{\tan\beta} $: in particular, the $H$ couplings to all fermions are suppressed if $\tan\beta \gg 1$. 
	
	The parameter space of the 2HDM (whatever the Type) is limited by  both  theoretical and experimental constraints.  
	The theoretical ones that have been imposed on the Higgs potential are as follows. 
	\begin{itemize}
		\item Perturbativity constraints imply that all quartic coefficients of the Higgs potential satisfy the condition $\mid\lambda_i\mid \leq 8\pi$ ($i=1,...5$)~\cite{Branco:2011iw}.
		\item Perturbative unitarity constraints require that $2\to2$ scattering processes involving Higgs and gauge bosons  remain unitary at high energy \cite{Kanemura:1993hm,Akeroyd:2000wc,Arhrib:2000is}.   
		\item Vacuum stability conditions require the Higgs potential to be bounded from below when the Higgs fields become large in any direction of the field space \cite{Deshpande:1977rw}.
	\end{itemize}
	We have used the public code 2HDMC-1.7.0 \cite{Eriksson:2009ws} to check the above theoretical constraints.
	
	We also take into account experimental constraints from Higgs analyses at lepton and hadron colliders (LEP, Tevatron and LHC) as well as EW Precision Observables (EWPOs). Limits from flavour physics observables are also considered.  Specifically, we have proceeded as follows. 
	
	\begin{itemize}
		\item Exclusion limits at $95\%$ Confidence Level (CL) from Higgs analyses at the aforementioned colliders  are implemented via HiggsBounds-5.9.0 \cite{Bechtle:2020pkv} and HiggsSignals-2.6.0 \cite{Bechtle:2020uwn}.
		
		\item We impose compatibility with the  EWPOs by requiring the computed $S, T$ and $U$ values  \cite{He:2001tp,Grimus:2008nb,Haber:2010bw} be within 2$\sigma$ of the SM fit of  \cite{ParticleDataGroup:2020ssz}, taking into account the full correlations among the three parameters.
		
		\item Constraints from $B$-physics observables are taken into account by using the public code SuperIso v4.1 \cite{Mahmoudi:2008tp}, in particular, we have used the following measurements:
		\begin{enumerate}
			
			\item[\textbullet]	${\rm BR}(\overline{B}\to X_s\gamma)|_{E_\gamma<1.6\mathrm{~GeV}}= \left(3.32\pm0.15\right)\times 10^{-4}$~\cite{HFLAV:2016hnz},
			\item[\textbullet]	${\rm BR}(B_s\to \mu^+\mu^-)_{\text{(LHCb)}}$ = $\left(3.09^{+0.46}_{-0.43}\right)\times 10^{-9}$~\cite{LHCb:2021awg,LHCb:2021vsc},
			\item[\textbullet]	${\rm BR}(B_s\to \mu^+\mu^-)_{\text{ (CMS)}}$=$\left(3.83^{+0.38}_{-0.36}\right)\times 10^{-9}$~\cite{CMS:2022mgd},
			\item[\textbullet]	${\rm BR}(B^0\to \mu^+\mu^-)_{\text{(LHCb)}}$=$\left(1.2^{+0.8}_{-0.7}\right)\times 10^{-10}$~\cite{LHCb:2021awg,LHCb:2021vsc}, 
			\item[\textbullet] ${\rm BR}(B^0\to \mu^+\mu^-)_{\text{(CMS)}}$=$\left(0.37^{+0.75}_{-0.67}\right)\times 10^{-10}$~\cite{CMS:2022mgd}.
		\end{enumerate}
	\end{itemize}

	\section{Numerical Results}
	As mentioned repeatedly, in this study, we focus on the inverted  mass hierarchy scenario, where the heaviest Higgs is identified as the observed 125 GeV at the LHC. We then scan randomly the 2HDM parameters over the  ranges of Tab.~\ref{tab0}.
	\begin{table}[!h]
		\begin{center}
			\resizebox{0.86\textwidth}{!}{
				\begin{tabular}{|c||c|c|c|c|c|c|} \hline
					Parameter&	$m_h$ (GeV) & $m_A$ (GeV) & $m_{H^\pm}$ (GeV)  & $\sin(\beta-\alpha)$ & $\tan\beta$ & $m_{12}^2$ (GeV$^2$) \\
					\hline
					Range & [10,~62] &  [62 ,~100]&  [96,~200] &    [-0.25, -0.05]&  [2,~25]&  $m_h^2\cos\beta \sin \beta$ \\
					\hline 
			\end{tabular}}
		\end{center}
		\caption {The input parameters of the  2HDM are tabulated and their scan ranges given. (Note that we have fixed $m_H = 125$ GeV.)}
		\label{tab0}
	\end{table}

\begin{figure}[h!]
		\centering	
		\includegraphics[scale=0.44]{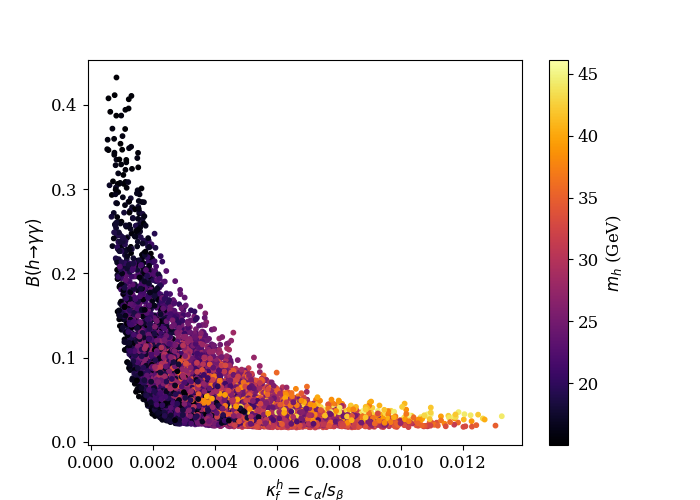}
		\includegraphics[scale=0.44]{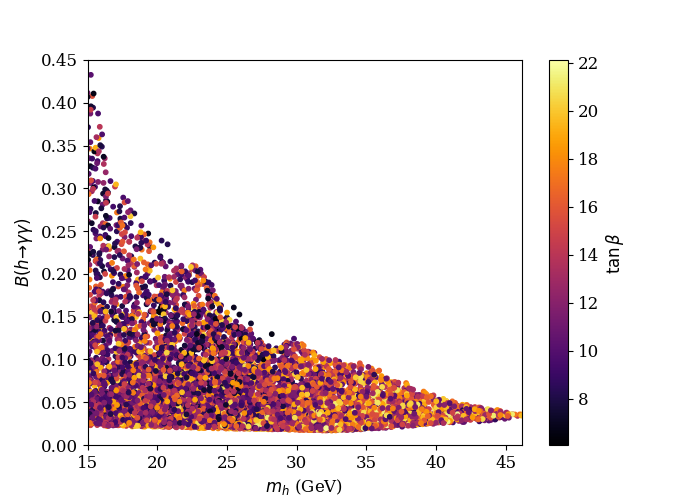}
		\caption{${\rm BR}(h \to \gamma \gamma)$ as a function of  $\kappa_f^{h}$ vs. $m_h$ (left panel) and as a function of $m_h$ vs. $\tan \beta$ (right panel) vs. $m_h$. }
		\label{fig0}
		\includegraphics[scale=0.44]{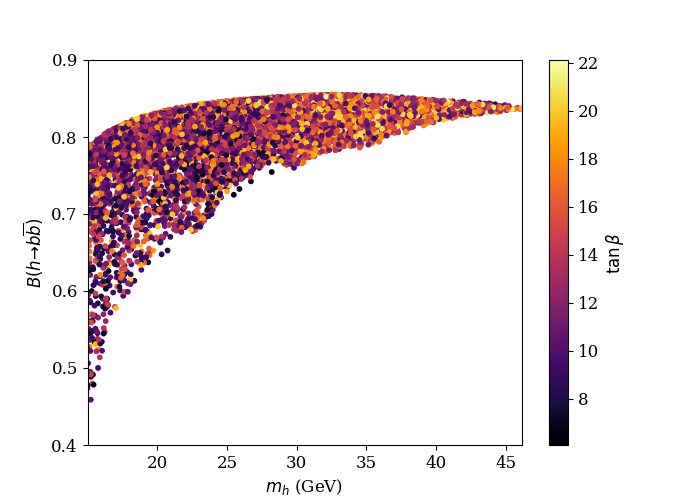}	
		\includegraphics[scale=0.44]{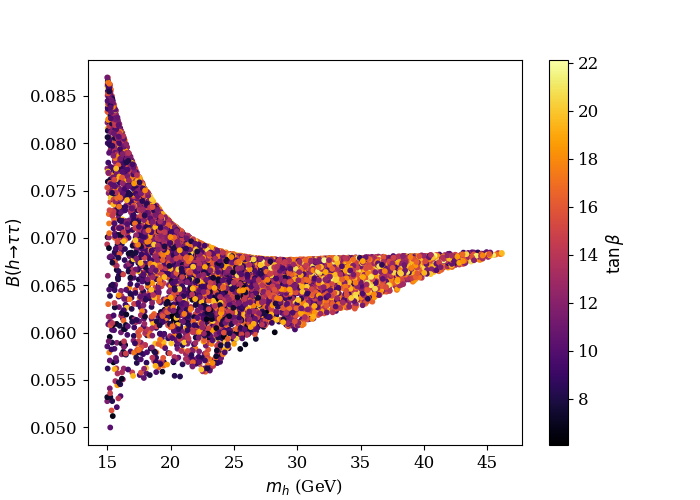}	
		\caption{${\rm BR}(h \to b\overline{b})$ (left panel) and ${\rm BR}(h \to \tau\tau)$ (right panel)  vs. $m_h$ and $\tan \beta$.}
		\label{fig1}
	\end{figure}

	Upon meeting the theoretical requirements and the constraints derived from past and ongoing experimental studies, described in section~\ref{section1}, we illustrate in Figs.~\ref{fig0} and \ref{fig1} the BRs  for different decay modes of the light Higgs: $h \to \gamma\gamma,~ b\overline{b}$ and $\tau \tau$. Away from the fermiophobic limit, where the light Higgs couplings to fermions $(\kappa_f^{h})$ are very suppressed ($c_\alpha / s_\beta \to 0$) and the di-photon decay of the light Higgs state can become significant, the total width of the $h$ state is clearly dominated by $h \to b\overline{b}$, over the entire 2HDM Type-I parameter space, with a BR of 85\%, followed by the decay into $\tau \tau$ with BR of order  $\sim 8\%$. The decay rate of $h \to b\overline{b}$ drops to 45\% when $\kappa_f^h$ approaches very small values ($\lesssim 0.001$) and $m_h$ is around 15 GeV. Furthermore, the partial width  $\Gamma(h \to b\overline{b})$ exhibits a  suppression near the $2m_b$ threshold and, therefore, $h \to \gamma \gamma$  can closely compete with $h \to b \overline{b}$ there.
	
	\begin{figure}[h!]
		\centering	
		\includegraphics[scale=0.44]{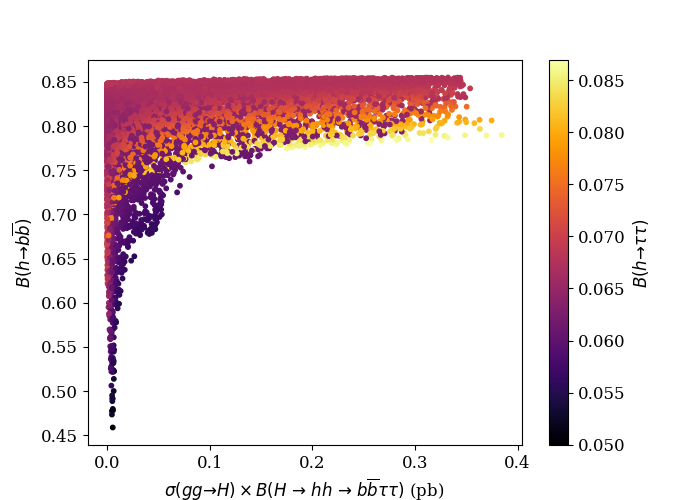}
		\includegraphics[scale=0.44]{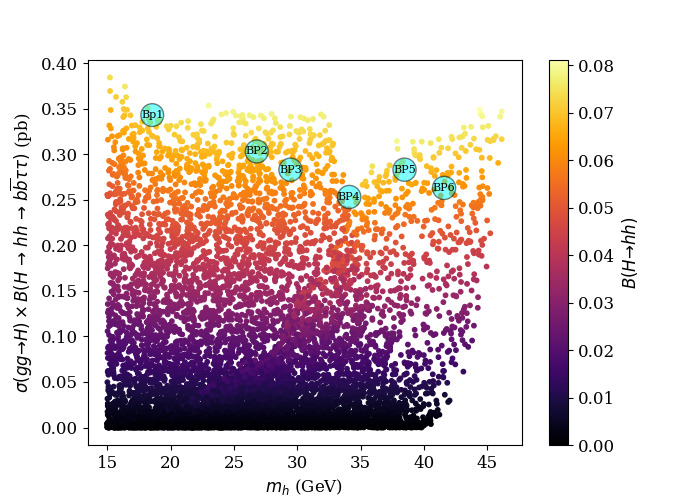}
		\caption{${\rm BR}(h \to b\overline{b})$ as a function of $\sigma(gg \to H \to hh \to b\overline{b}\tau\tau)$ vs. ${\rm BR}(h \to \tau \tau)$ (left panel)  and $\sigma(gg \to H \to hh \to b\overline{b}\tau\tau)$ as a function of $m_h$ vs. ${\rm BR}(H \to hh)$ (right panel).}
		\label{fig2}
	\end{figure}
	
	Fig.~\ref{fig2} shows the overall cross-section of the process $pp\to H \to hh \to b\overline{b}\tau \tau$, wherein the main channel for producing the SM-like Higgs boson $H$ is the gluon-gluon fusion 
	mechanism. Here, ${\rm BR}(h \to b\overline{b}), ~{\rm BR}(h\to \tau \tau)$ and ${\rm BR}(H \to hh )$ are also displayed. Since the decay width of the SM-like Higgs  ($\Gamma_H$) is of the order of a few MeV ($\Gamma_H^{\text{ATLAS}} = 4.6^{+2.6}_{-2.5}$ MeV~\cite{ATLAS:2023dnm}  and $\Gamma_H^{\text{CMS}}=3.2^{+2.4}_{-1.7}$ MeV~\cite{CMS:2022ley}), one can assume the Narrow Width Approximation (NWA)\footnote{The same applies to the $h$ state, which would be even narrower.} and then write the complete production times decay cross-section as follows:
	\begin{eqnarray}
	\sigma_{b\overline{b}\tau\tau} = \sigma(gg \to H)\times {\rm BR}(H \to hh) \times {\rm BR}(h \to b\overline{b})\times {\rm BR}(h\to \tau \tau), 
	\end{eqnarray}
	where $\sigma(gg \to H)$ is the $H$ production rate computed (here) at Leading Order (LO) using \texttt{SusHi 1.7.0}~\cite{Harlander:2012pb,Harlander:2016hcx,Harlander:2002wh} at the center-of-mass energy of 13 TeV. (The N$^3$LO QCD corrections will be considered through a $K$-factor~\cite{Anastasiou:2015vya,Cepeda:2019klc}, see below.) As shown previously in Eq.~(\ref{tri-coupling}), the trilinear Higgs coupling $Hhh$ is proportional to $\cos(\beta-\alpha)$. In the inverse mass hierarchy configuration of the 2HDM Type-I, current data drive $\cos(\beta-\alpha)$ to closely approach 1\footnote{The allowed values of the SM-like Higgs couplings to vector bosons, $\kappa^H_V =\cos(\beta-\alpha)$, $V=W^\pm,Z$, lie very close to 1, with a deviation of 2\% from the SM values.} and thus ${\rm BR}(H \to hh)$ is not suppressed. However, it is observed that the BR of the SM-like Higgs boson ($H$) decaying into a pair of light Higgs bosons ($hh$) remains below 10\% when the decay channel becomes kinematically accessible. This limitation arises from the precise measurements of the Higgs boson couplings, which impose stringent constraints on exotic decays of the Higgs boson, i.e., into  BSM particles and/or undetected final states (invisible, for short). As mentioned, the ATLAS and CMS collaborations have set, respectively, an upper limit of 12\% \cite{ATLAS:2022vkf} and 16\% \cite{CMS:2022dwd} on ${\rm BR}(H \to {\rm invisible})$ at 95\% CL, using LHC Run 2 data. Recently, the ATLAS collaboration also performed a combination of Run 1 and 2 direct searches for invisible Higgs decays, where several production modes of the SM-like Higgs boson are considered~\cite{ATLAS:2023tkt}.  An upper bound of 10.7\% (7.7\%) on ${\rm BR}(H \to {\rm invisible})$ at the 95\% CL has been observed (expected). The CMS collaboration has also lately presented the combination of a search for $H \to {\rm invisible}$ with the SM-like Higgs state produced in association with a top-antitop pair  (i.e., $pp\to t\overline{t}H$) or a vector boson (i.e., $pp\to VH$) using  $138~\text{fb}^{-1}$ of data from both Run 1 and 2: the combined upper limit on ${\rm BR}(H \to {\rm invisible})$ is 15\% at 95 CL~\cite{CMS:2023sdw}. Compatibly with these constraints, we find that the cross-section for the process $gg \to H \to hh \to b\overline{b}\tau\tau$ reaches its maximum value of 0.4 pb when  ${\rm BR}(h \to b\overline{b})$, ${\rm BR}(h \to \tau \tau)$ and ${\rm BR}(H\to hh)$  are at their maximum values. Over the parameter region allowing for this, we have marked several Benchmark Points (BPs) amenable to MC simulation, which are listed in Tab.~\ref{tab1}.
	
	An additional scan is then performed with fixed values of $m_{H^\pm} = 165.58~\text{GeV}$, $m_{A} = 98.9~\text{GeV}$, $\sin(\beta-\alpha) = -0.10$, $m_{12}^2 = 154~\text{GeV}^2$ while $m_h$ and $\tan \beta$ are varied randomly as shown in Tab.~\ref{tab0}.
	\begin{figure}[h!]
		\centering		
		\includegraphics[scale=0.34]{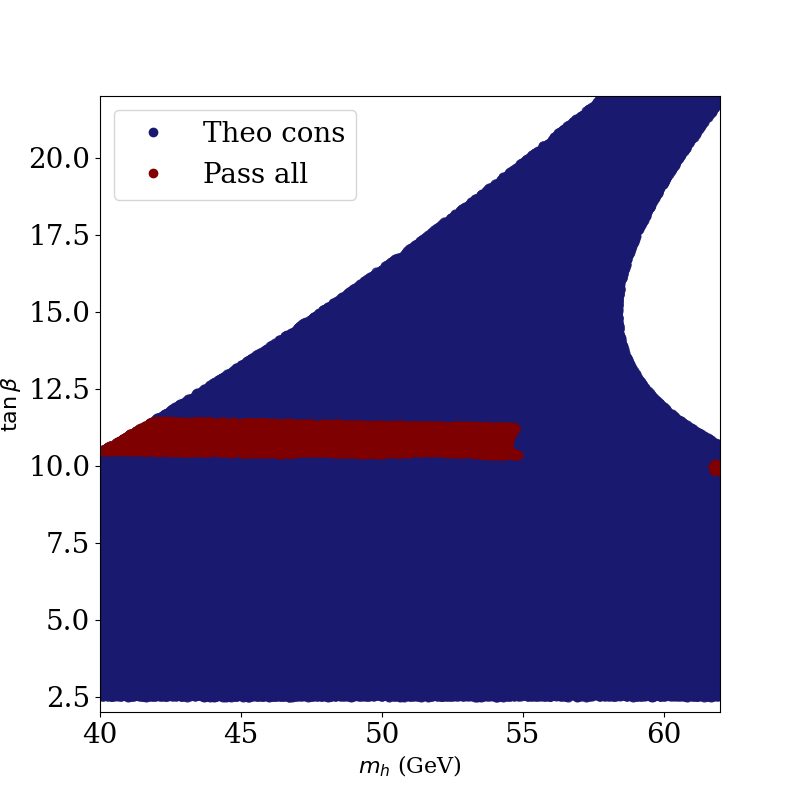}
		\includegraphics[scale=0.34]{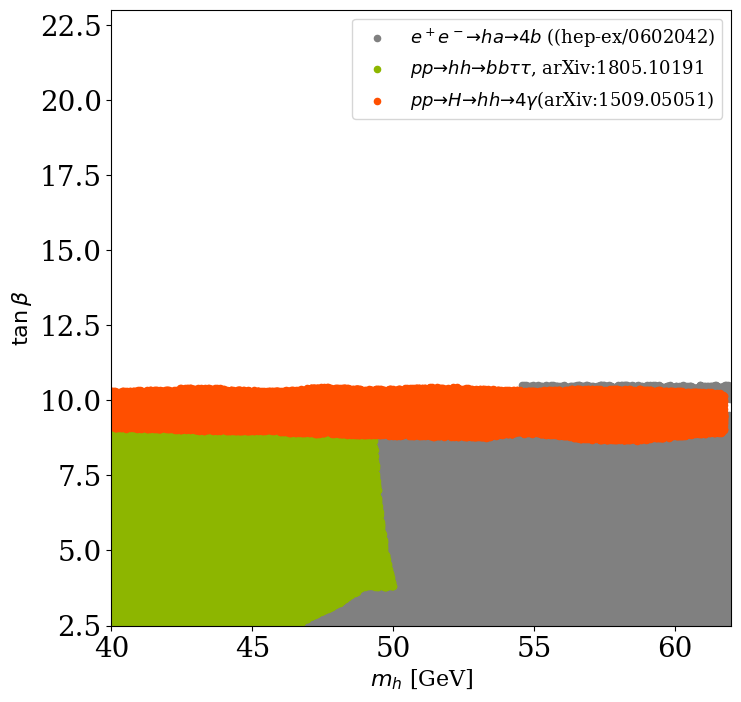}
		\caption{Allowed (left panel)  and excluded (right panel) parameter space over the ($m_h,~\tan \beta$) plane. Here, $m_{H^\pm} = 165.58~\text{GeV},~m_{A} = 98.9~\text{GeV},~\sin(\beta-\alpha) = -0.10,~m_{12}^2 = 154~\text{GeV}^2$.}
		\label{fig3}
	\end{figure}
	We show in the left panel of Fig.~\ref{fig3} the allowed region resulting from passing the theoretical constraints, indicated by blue colour. The points that meet the theoretical requirements and also align with experimental data are highlighted in red colour.
	To gain a better understanding of this restricted parameter space, which is limited to a narrow range of values of $\tan \beta$, specifically between 9 and 11.50, we illustrate in the right panel of Fig.~\ref{fig3} the excluded regions and the corresponding searches. For values of $\tan \beta$ below 9, the CMS search for exotic decays in the $b\overline{b}\tau\tau$~\cite{CMS:2018zvv} final state has excluded the mass range where $m_h < 50~\text{GeV}$ whereas the region where $m_h > 50~\text{GeV}$ is excluded by the LEP search for $e^+ e^- \to hA \to b\overline{b}b\overline{b}$~\cite{ALEPH:2006tnd}. Additionally, an ATLAS search  for events with at least three photons (i.e., $3\gamma$) targeting  the process $pp \to H \to hh \to 4\gamma$~\cite{ATLAS:2015rsn} has led to the exclusion of the parameter space with $\tan \beta > 9$ and $40~\text{GeV}<m_h<62.5~\text{GeV}$. 
	
	\begin{figure}[h!]
		\centering	
		\includegraphics[scale=0.44]{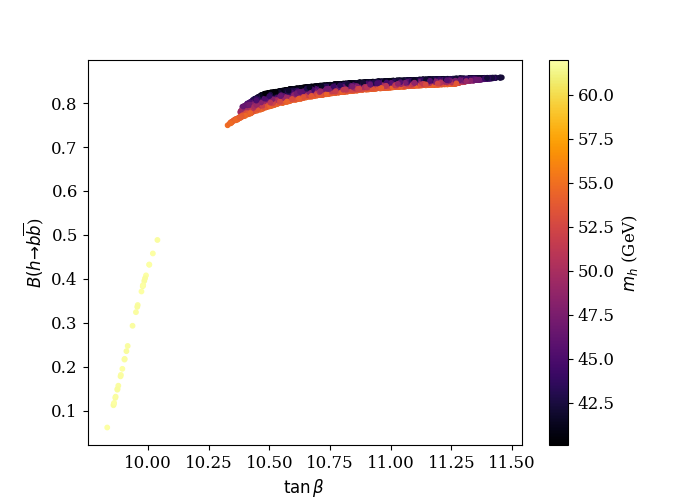}	
		\includegraphics[scale=0.44]{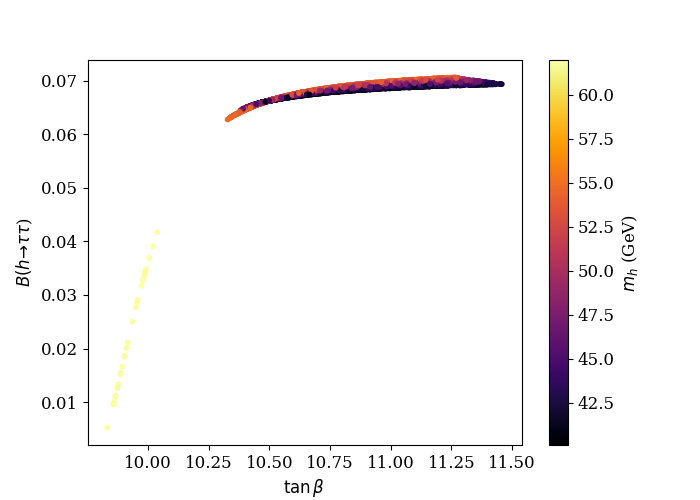}	
		\caption{${\rm BR}(h \to b\overline{b})$ (left panel) and ${\rm BR}(h \to \tau\tau)$ (right panel) as a function of $\tan \beta$ vs. $m_h$. Here, $m_{H^\pm} = 165.58~\text{GeV},~m_{A} = 98.9~\text{GeV},~\sin(\beta-\alpha) = -0.10,~m_{12}^2 = 154~\text{GeV}^2$.}
		\label{fig4-a}
	\end{figure}
	
	We show in Fig.~\ref{fig4-a} the different BRs within the described parameter space. Similarly to Fig.~\ref{fig0}, the dominant channel within the parameter space where  $m_h<60~\text{GeV}$ is $h \to b\overline{b}$, with a BR that can reach up to 85\%. The second prominent channel is $h \to \tau \tau$ with a BR that goes up to 8\%. Another interesting observation is that the fermionic decay rates of  $h \to b\overline{b}$ and $\tau\tau$ remain relatively unaffected by $\tan \beta$ variations. For large $\tan \beta$, the Yukawa couplings\footnote{Notice that $\kappa^f_h = \sin(\beta-\alpha) + \cot \beta \cos(\beta-\alpha) \approx \sin(\beta-\alpha)$ for large $\tan \beta$, with $\sin(\beta-\alpha) = -0.10$.} become suppressed, leading the BRs of the light Higgs state decaying into fermions to become fairly independent on the values of $\tan \beta$ within the considered parameter space. 
	
	\begin{figure}[h!]	
		\centering
		\includegraphics[scale=0.44]{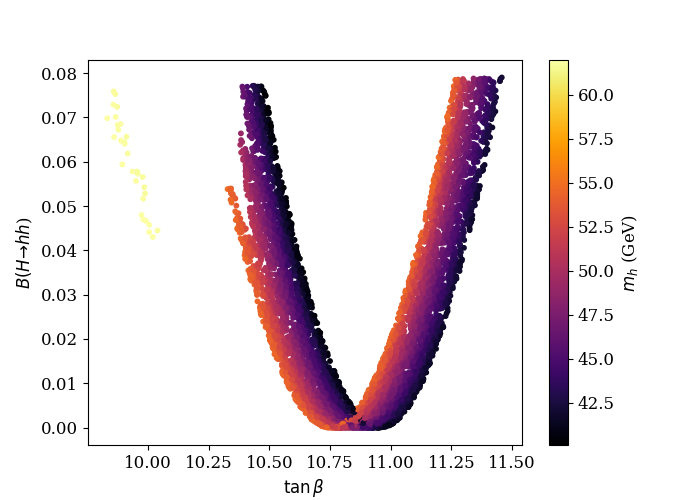}
		\includegraphics[scale=0.44]{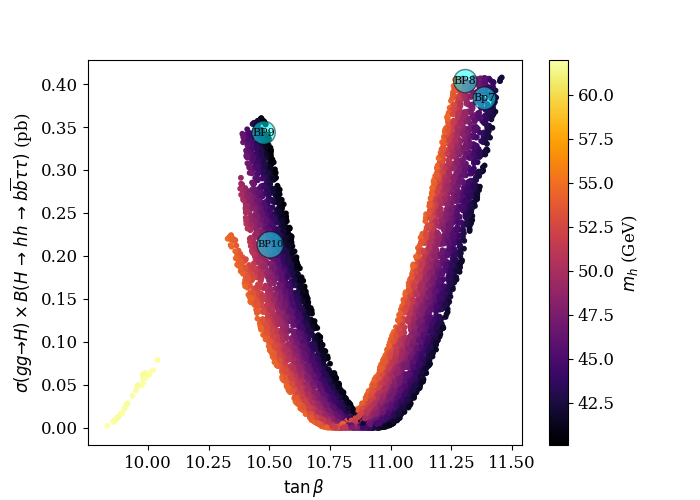}	
		\caption{${\rm BR}(h \to hh)$ (left panel) and $\sigma(gg \to H)\times {\rm BR}(H \to hh \to b\overline{b}\tau \tau)$ (right panel) as a function of $\tan \beta$ vs. $m_h$. }
		\label{fig4-b}
	\end{figure}
	
	In Fig.~\ref{fig4-b}, we present the production cross-section of the SM-like Higgs state $H$ times its decay BR into $b\overline{b}\tau\tau$ via $hh$, i.e., $\sigma(gg\to H)\times {\rm BR}(H \to hh \to b\overline{b}\tau\tau)$, within the specified parameter space. The maximum value of this product is observed to be 0.4 pb, which occurs, again, when ${\rm BR}(H \to hh)$, ${\rm BR}(h \to \tau \tau)$ and ${\rm BR}(h \to b\overline{b})$ all reach their maximum values within the considered parameter space. 
	Similarly to the previous scenario, we have selected a few BPs to perform a MC simulation. These BPs are carefully chosen to cover a range of interesting scenarios and to explore various aspects of the model, see Tab.~\ref{tab1}.
	
	\section{Feasibility Study at the LHC Run 3}
	
	In our MC analysis, we focus on the $b\bar b\tau\tau $ final state, where both $\tau$ leptons decay into either an electron or a muon, along with their respective neutrinos. To distinguish between the decays of the $\tau$ particles, we use the short-hand notations $\tau_e$ and $\tau_\mu$ to represent the channels $\tau \to e \bar{\nu}_e \nu_\tau$ and $\tau \to \mu \bar{\nu}_\mu \nu_\tau$, respectively. 
	The final states $\tau_e\tau_e$ and $\tau_\mu \tau_\mu$ are neglected in order to suppress the significant contamination from Drell-Yan (DY) background events. In Fig.~7, we present the Feynman diagram of the signal process, where effective vertices are considered for the gluon-gluon-Higgs coupling and for leptonic tau decay~\cite{Hagiwara:2012vz}. 
	
	\begin{figure}[h!]
		\centering	
		\includegraphics[scale=0.55]{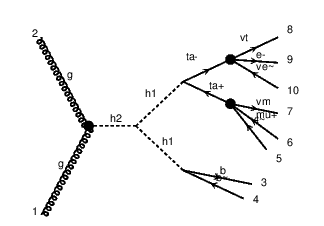}
		\caption{Feynman diagram for the process $gg \to H \to hh \to b\overline{b}\tau_e \tau_\mu$. Here, $h_2\equiv H$ and $h_1 \equiv h$.}
		\label{fig5}
	\end{figure}
	
	We use \texttt{MadGraph-v.3.4.2}~\cite{Alwall:2014hca} to generate parton-level events of both signal and background processes. To account for the $\tau$ decays, we use the {\tt TauDecay} library~\cite{Hagiwara:2012vz}\footnote{The {\tt TauDecay} package has been added to the UFO file of the 2HDM Type-I.}. The QCD corrections are taken into account by using $K$-factors for (what we will prove to be) the two main background processes $ Z(\to \tau_e \tau_\mu)b\overline{b}$~\cite{FebresCordero:2009xzo} and $t\overline{t}$ \cite{SM:2010nsa}: specifically, $K=1.4$ for both $b\overline{b}Z$ and $t\overline{t}$. The $K$-factor for the signal production process $gg \to H$ is taken as $K=2.5$, which includes the N$^3$LO QCD corrections of Refs.  \cite{Anastasiou:2015vya,Cepeda:2019klc}. We then pass the events to \texttt{PYTHIA8}~\cite{Sjostrand:2006za} for parton showering, fragmentation/hadronisation and heavy flavour decays. Then, we use \texttt{Delphes-3.5.0}~\cite{deFavereau:2013fsa} with a standard CMS card to simulate the detector response. Finally, we employ \texttt{MadAnalysis 5}~\cite{Conte:2012fm} to apply cuts and to conduct the kinematic analysis.
	As mentioned already, 
	the BPs and the corresponding parameter values used to generate MC samples of events for the signal process are given  in Tab.~\ref{tab1}, where, as usual, the collision energy at the proton-proton level is assumed to be 13 TeV.
	
	\begin{table}[!h]
		\begin{center}
			\resizebox{0.99\textwidth}{!}{
				\begin{tabular}{|c|c|c|c|c|c||c|c|c|c|} \hline
					BP  & $m_h$ (GeV) &  $m_a$ (GeV)  & $m_H^\pm$ (GeV) & $\sin(\beta-\alpha)$ & $\tan \beta$ & $\sigma_{b\overline{b}\tau\tau}$ (pb)& ${\rm BR}(H \to hh)$ & ${\rm BR}(h \to b\overline{b})$ & ${\rm BR}(h \to \tau\tau)$\\
					\hline
					BP1 & 17.67 & 73.70& 184.51&-0.053 & 19.68 & 0.34 & 0.07 & 0.81 & 0.076\\
					\hline
					BP2& 25.9 & 80.61 &171.88 & -0.064 & 16.71 & 0.3 & 0.068 & 0.84 & 0.068\\
					\hline
					BP3 & 28.56 & 94.46 & 155.45 & -0.11& 9.09 & 0.28 & 0.065 & 0.85 & 0.067\\
					\hline
					BP4 & 33.20 & 88.29& 99.75&-0.076 & 14.42 & 0.25 & 0.058 & 0.85 & 0.067 \\
					\hline
					BP5 & 37.56 & 88.88 & 188.64 & -0.064 & 16.45 & 0.28 & 0.072 & 0.8 & 0.064\\
					\hline
					BP6 & 40.68 & 88.37 & 144.39 & -0.054 & 19.37 & 0.26 & 0.063 & 0.82 & 0.066\\
					\hline
					BP7 & 47.27 & 98.91 & 165.58 & -0.10 & 11.34 & 0.38 & 0.074 & 0.85 & 0.074\\
					\hline
					BP8 & 54.03 & 98.91 & 165.58 & -0.10 & 11.28 & 0.40 & 0.083 & 0.84 & 0.07\\
					\hline
					BP9 & 43.44 & 98.91 & 165.58 & -0.10 & 10.43&  0.34 &0.077 & 0.80 & 0.065\\				
					\hline
					BP10 & 49.39 & 98.91 & 165.58 & -0.10 &	10.41 & 0.21 & 0.056 & 0.78 & 0.065 \\	
					\hline  
			\end{tabular}}
		\end{center}
		\caption {The cross-sections  $\sigma_{b\overline{b}\tau\tau}\equiv\sigma(gg\to H)$ $\times$ BR($H\to hh)$ $\times$ BR$(h\to b\bar b)$ $\times$ BR$(h\to\tau\tau$) for our BPs are given for the collision energy $\sqrt{s}=13$ TeV alongside the BRs for the decay channels $H\to hh$, $h\to b\bar b$ and $\tau\tau$. The unit of all masses is GeV. Here, $m_H = 125$ GeV, and in the values of 
  cross-section $\sigma_{b\bar{b}\tau\tau}$ the $K$-factor has been taken into consideration.}
		\label{tab1}
	\end{table}
	
	In order to generate signal and background events efficiently, we apply the following kinematic cuts at {parton level}:
	\begin{equation*}
	p_T(b) > 10~\text{GeV},~p_T(l)>5~\text{GeV},~E_T^{\rm miss} > 5~\text{GeV}, ~|\eta(b,l)|<2.5,~\Delta R(ll,bl,bb) > 0.3,~H_T\footnote{\rm Where $H_T$ is computed over all partons.}<70~\text{GeV}.
	\end{equation*}
	
	The LO cross-sections of all background processes considered in our initial MC analysis are given in Tab. \ref{tab2}. As intimated, it is found that the dominant background processes arise from top pair production ($t\overline{t}$) and $Z(\to \tau_e \tau_\mu)b\overline{b} $, so that we will simulate only these two processes in our final (detector level) analysis. 
	\begin{table}[!h]
		\begin{center}
			\resizebox{0.44\textwidth}{!}{
				\begin{tabular}{|c|c|c|c|} \hline
					Background process  &  $\sigma$ (pb)  \\ 
					\hline
					$pp \to Z(\to b\overline{b}) Z(\to  ll)$, $l = (e,~\mu,~\tau_{e,~\mu})$&  0.009 pb \\
					\hline
					$pp \to  Z(\to ll)b \overline{b}$, $l = (e,~\mu,~\tau_{e,~\mu})$&  6.1 pb \\
					\hline
					$pp \to Z(\to b \overline{b}) ll$, $l = (e,~\mu,~\tau_{e,~\mu})$&  0.015 pb \\
					\hline
					$pp \to Z {W^\pm} j,~ {W^\pm} \to l \nu_l~(l = e,~\mu,\tau_{e,~\mu})$  & 0.0051 pb	\\
					\hline
					$pp \to t\overline{t} \to e^\pm \mu^\mp  b\overline{b}+E_T^{\rm miss}$ &0.28 pb \\
					\hline 
			\end{tabular}}
		\end{center}
		\caption {The LO background cross-sections are given for the collision $\sqrt{s}=13$ TeV. The $K$-factors  mentioned in the text has not been included here but will be used to compute the final significances, albeit limitedly to the second and last processes (i.e., the two dominant noises).
		}
		\label{tab2}
	\end{table}
	
	\subsection{Parton Level Analysis}
	
\begin{figure}[h!]
		\centering	
		\includegraphics[scale=0.4]{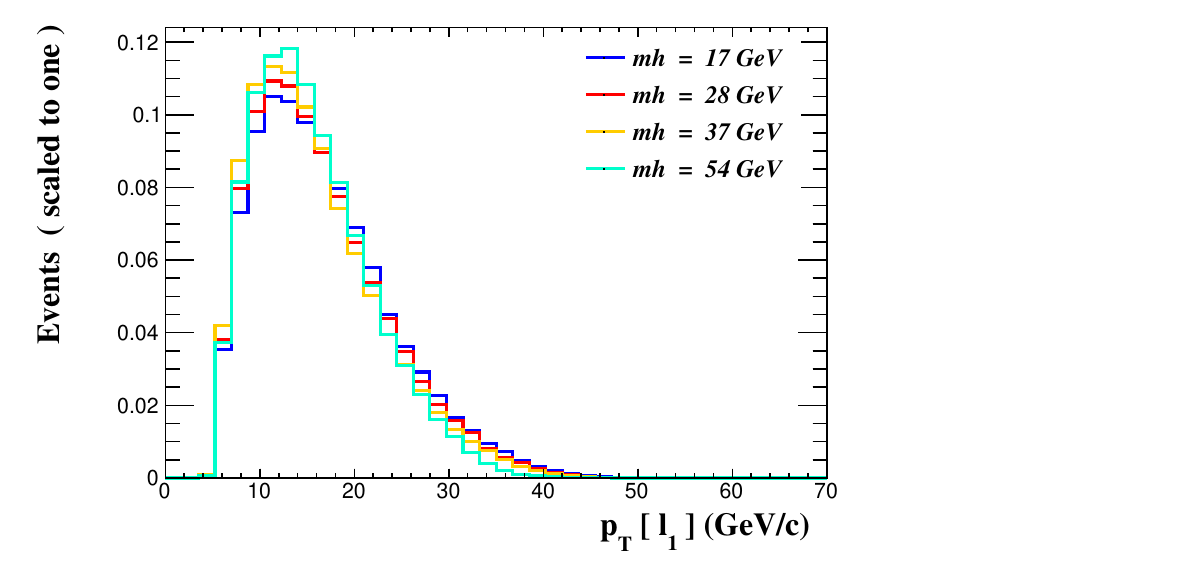}	
		\includegraphics[scale=0.4]{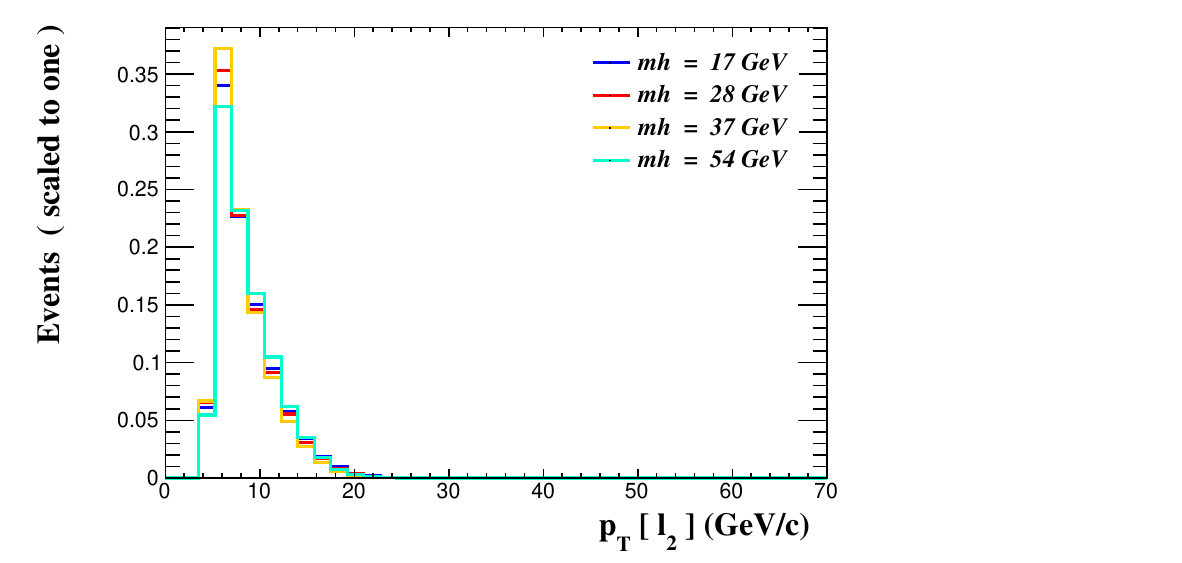}	
	    \includegraphics[scale=0.4]{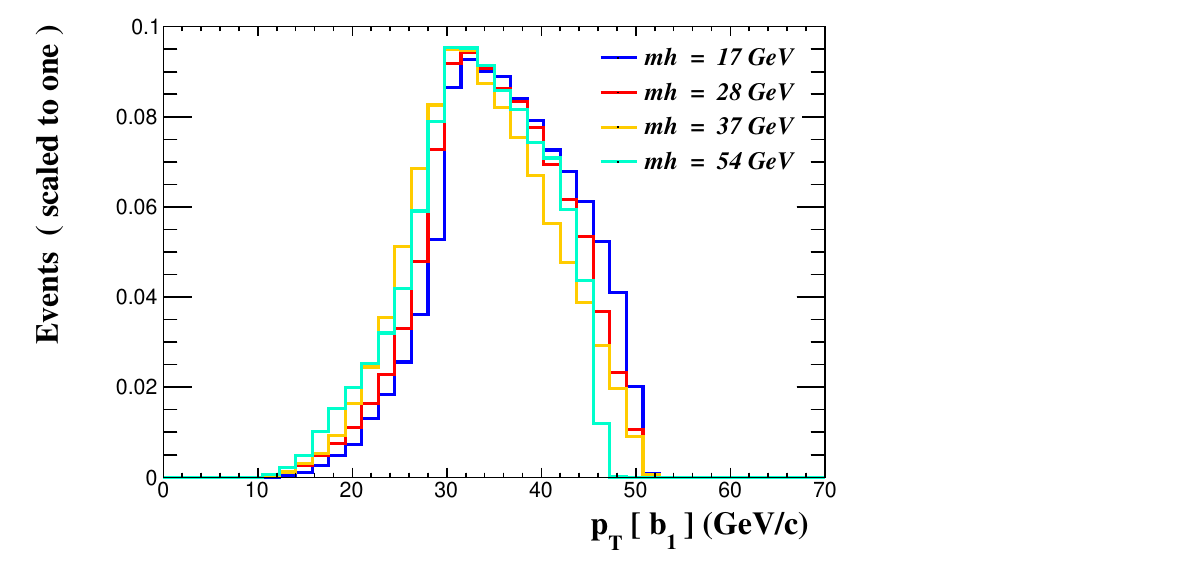}	
	    \includegraphics[scale=0.4]{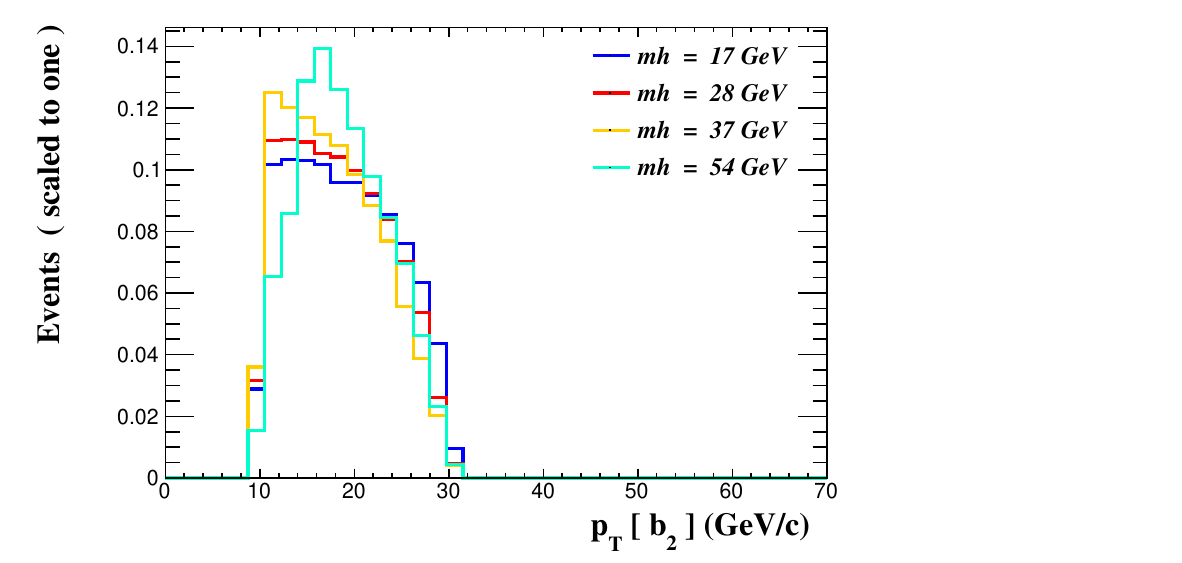}	
		\caption{The $p_T$ distributions of the leading (subleading) lepton (top panel) and leading (subleading) $b$-(anti)quark (bottom panel) for different signal BPs at are shown at parton level, see Tab.~\ref{tab2}.}
		\label{figg5}
	\end{figure}

	In Fig.~\ref{figg5}, the $p_T$ distributions of the leading and subleading leptons as well as of the $b$-(anti)quarks for the signal are displayed. These spectra provide insights into the transverse momentum of final-state particles involved in the signal process, in order to guide our final analysis at detector level.  
	
	On the one hand, it is observed that the leading and subleading leptons originating from the light Higgs decay (recall that $m_h < 62.5~\text{GeV}$) tend to have lower transverse momentum, indicating that they are relatively softer when compared with the leading (subleading) $b$-(anti)quarks from the same light Higgs state (owing to neutrinos carrying away energy). In particular, the peak value of $p_T$ of the subleading lepton is less than 10 GeV, which means a stiff (detector level) cut on $p_T(l)$ may lead to a severe loss of signal events.
	On the other hand, since the $b$-tagging efficiency decreases with $p_T$, then one should expect that a stiff (detector level) cut on, especially,  the subleading $b$-jet can also lead to a severe loss of signal events. 
	In short, these factors call for a dedicated trigger choice, that we will illustrate below.

\begin{figure}[h!]
		\centering		
		\includegraphics[scale=0.4]{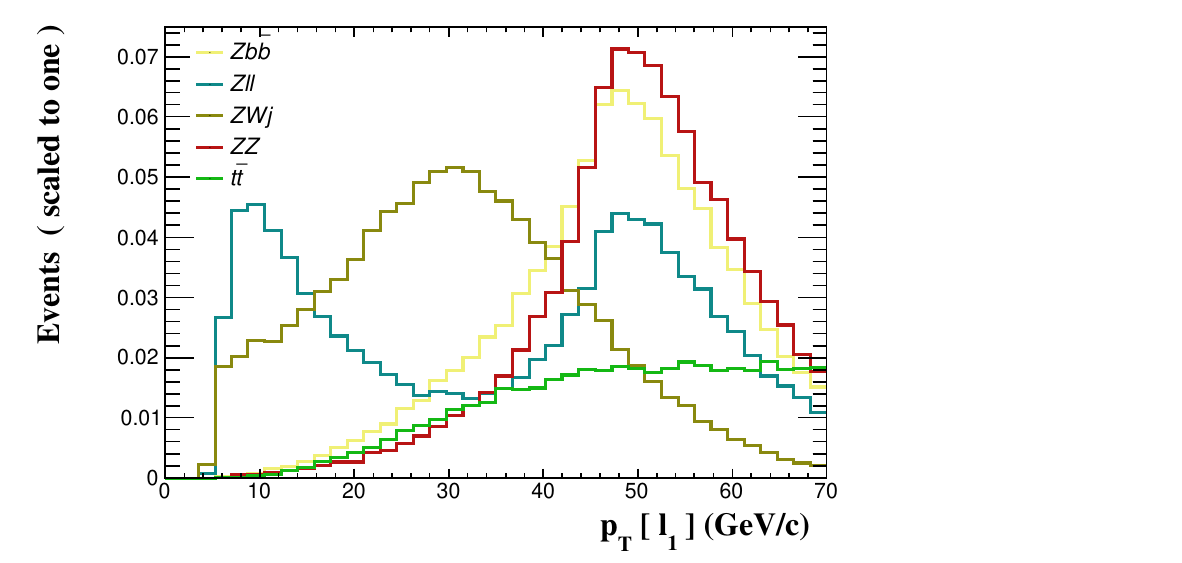}	
		\includegraphics[scale=0.4]{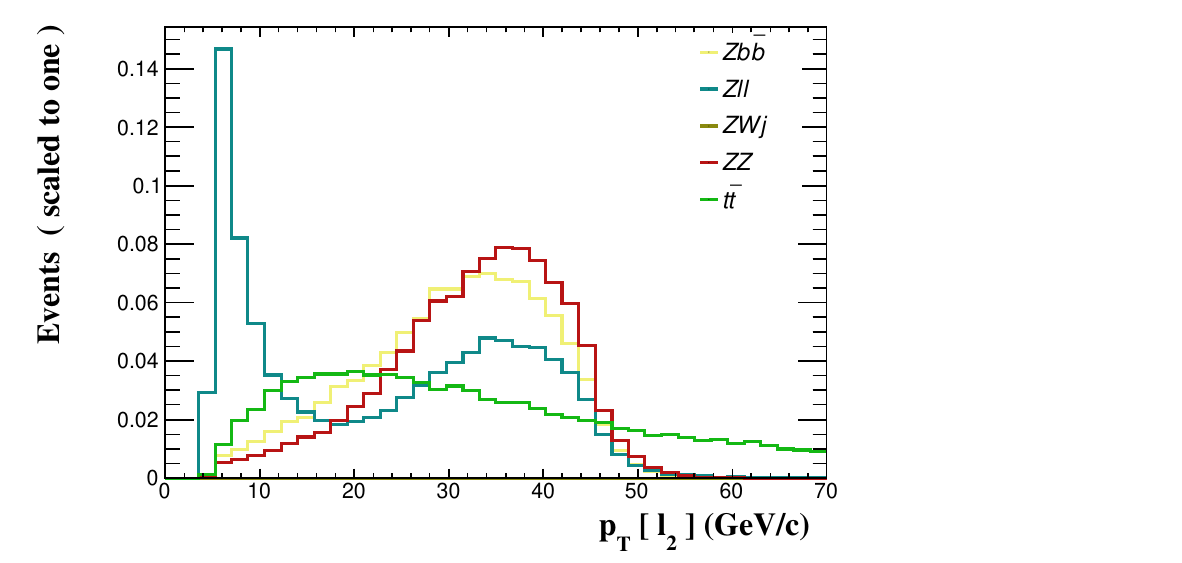}		\\
		\includegraphics[scale=0.4]{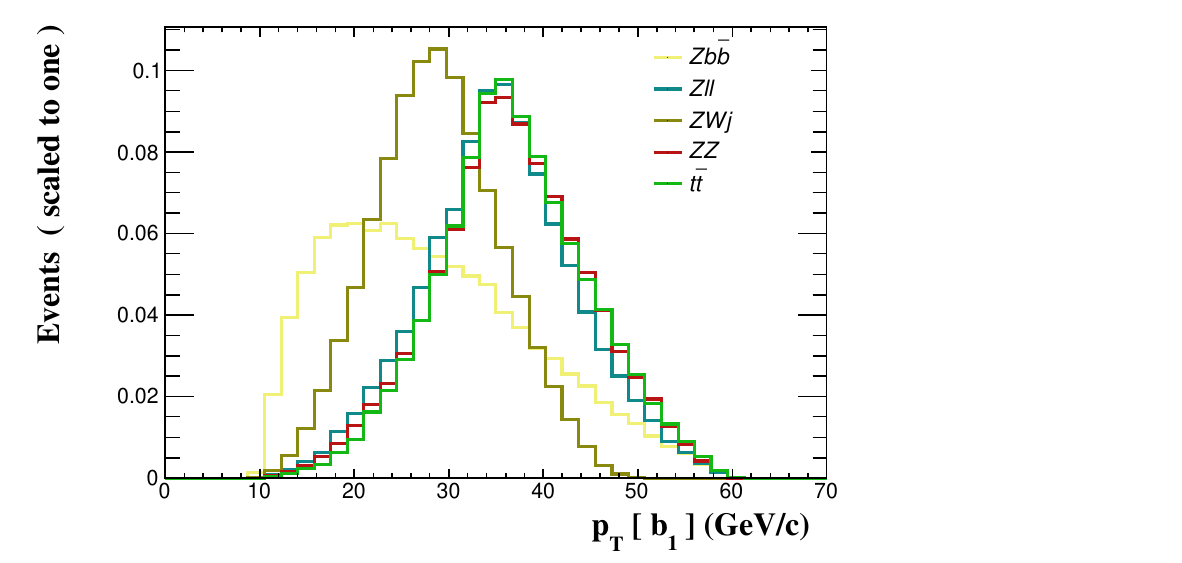}	
		\includegraphics[scale=0.4]{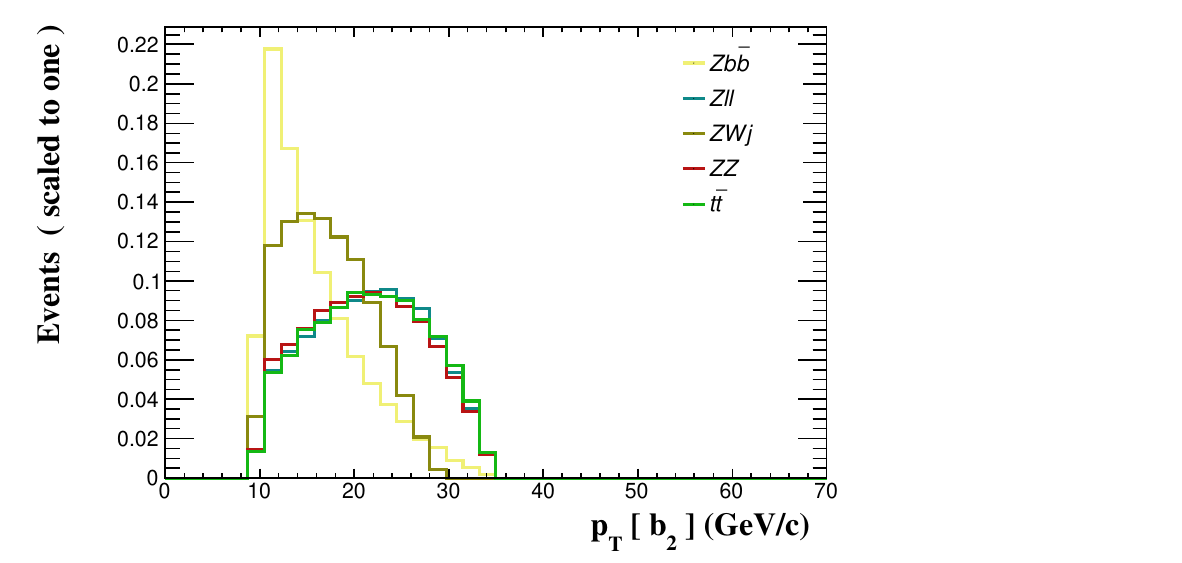}	
		\caption{The $p_T$ distributions of the leading (subleading) lepton and $b$-(anti)quark of different background processes are shown at parton level, see Tab.~\ref{tab2}.}
		\label{fig6}
	\end{figure}

	In Fig. \ref{fig6}, we show the leading (subleading) lepton and $b$-(anti)quark of different background processes at parton level. It is observed that the subleading lepton from $Z(\to b\bar{b})\tau_e\tau_\mu$ can be very soft, just like the subleading lepton from the signal events. Similarly, the subleading $b$-(anti)quark from $Z(\to\tau_e\tau_\mu)b\bar b$ can be very soft, which can then mimic the subleading $b$-(anti)quark of the signal events. Thus, this renders even more important the accurate choice of a dedicated trigger for this analysis. 
	However, as intimated, from Tab. \ref{tab2}, it is clear that the dominant background processes are $pp \to Z (\to\tau_e\tau_\mu)b \bar b$ and $pp \to t \bar{t}$. Therefore, in order to discover the signal process, the key task is to ultimately suppress these two backgrounds. Armed with the knowledge acquired at the parton level, we will investigate below how to suppress these two types of  noise  most effectively at the detector level. 
	
	\subsection{Detector Level Analysis}
	
\begin{figure}[h!]
		\centering	
		\resizebox{0.32\textwidth}{!}{
			\includegraphics{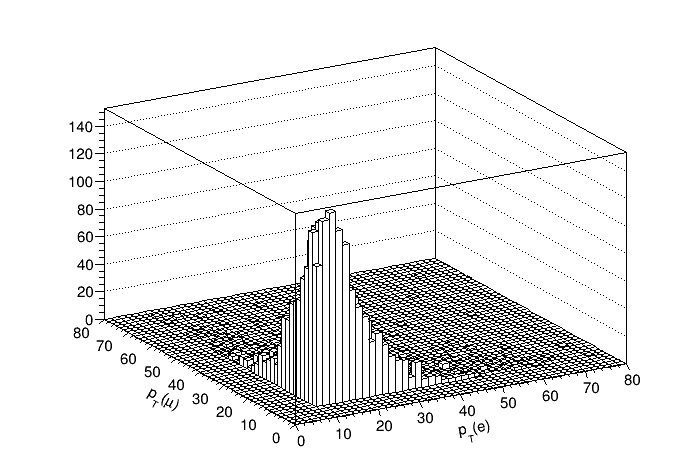}}
		\resizebox{0.32\textwidth}{!}{
			\includegraphics{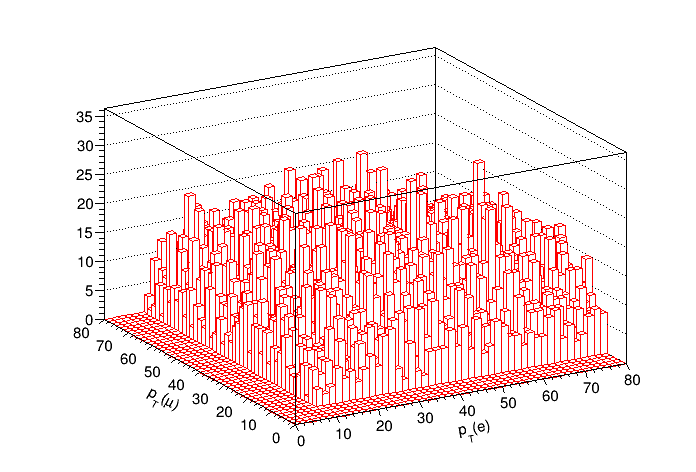}}
		\resizebox{0.32\textwidth}{!}{
			\includegraphics{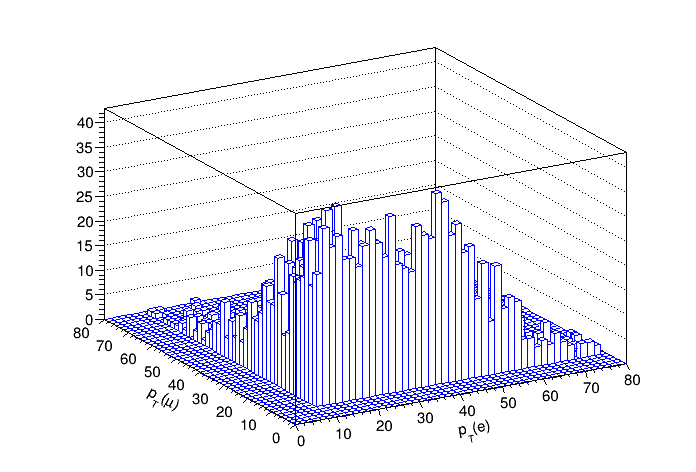}}
		\caption{Correlations between the $p_T$ of the leading and subleading leptons for signal  (black), $t\overline{t}$ (red) and $b\bar bZ(\to\tau_e\tau_\mu)$  (blue) are shown at detector level.
		}
		\label{fig7}
	\end{figure}

	We only consider events with two $b$-jets and two leptons of different flavour with opposite charges (i.e., $e^\mp\mu^\pm$) in the final state. Given that both (leading and subleading) leptons in the signal events are soft, as illustrated in Fig.~\ref{fig7}, it will be necessary to suitably adjust the transverse momentum ($p_T$) thresholds of the leading electron ($e$) and subleading muon ($\mu$). By reducing these $p_T$ thresholds, one can include more signal events and then increase the acceptance so to mitigate the potential significant loss of signal events due to the softness of the leptons, especially the subleading one. Consequently, the  trigger choice could lead to a more comprehensive view of the final state and potentially improve the overall sensitivity of the analysis. 
	
Events featuring tau decays into leptons ($\tau_e\tau_\mu$) are collected by electron and muon triggers. The CMS cross trigger~\cite{CMS:2016ngn,CMS:2018jrd} relies on the presence of both an electron ($e$) and a muon ($\mu$), where the leading lepton has a $p_T$ threshold of 23 GeV and the subleading one has a $p_T> 12~\text{GeV}$ for an electron or 8 GeV for a muon~\cite{CMS:2018zvv,CMS:2023ryd}. This trigger is limited by $p_T$ thresholds in the Level-1 (L1) trigger selection due to the limited L1 bandwidth available. The double electron trigger is similarly limited. Recently, in Run 3, CMS has overcome this limitation for electron pairs that are close together, for example from B meson decays, by placing tighter topological selection on the two electrons L1 objects. This borrows from established approaches already used in double muon triggers searching for low $p_T$ muons from B meson decays.  By applying similar topological approaches to an electron muon trigger, it would seem possible to develop a reasonably efficient L1 selection targeting ~10 GeV close by electron muon pairs with a rate in the $\sim 2-5~\text{kHz}$ range at nominal Run 3 luminosities.  A L1 selection with such a bandwidth is feasible to be deployed in Run 3 if there is sufficient will on the experiment's side to do so and would be easily achievable at the HL-LHC.  At the high level trigger (HLT) selection stage, the rate should be manageable and if not, scouting and parking techniques~\cite{Mukherjee:2019anz,Bainbridge:2020pgi}  can be used if the rate of the selection is too much to be included in the standard physics stream.  Electrons and muons from $\tau$ decays are then selected after satisfying the following requirements:
	\begin{equation*}
	p_T^e > 10~\text{GeV},~ p_T^\mu > 8~\text{GeV},~|\eta^e|<2.4,~|\eta^\mu|<2.4.
	\end{equation*}
	Lepton candidates must also satisfy isolation criteria, which ensure that the amount of activity in a cone of radius $R = 0.4$ centered on the muon (electron) direction is smaller than 20\% (15\%) of $p_T^\mu~(p_T^e)$. For the jets candidates, we employ the anti-$k_T$ algorithm~\cite{Cacciari:2008gp} to cluster detector-level objects with a jet radius parameter $\Delta R$ = 0.4 and $p_{T,j}^{\rm min} = 10$ GeV (for both light- and $b$-jets)\footnote{Note that our results are rather stable against a different choice of jet clustering algorithm, like the Cambridge-Aachen one \cite{Dokshitzer:1997in,Wobisch:1998wt}.}. All events are required to have two $b$-tagged jets with $p_T(b_1/b_2)>10~\text{GeV}$ (where $p_T(b_{1~(2)})$ represents the $b$-jet with highest (lowest) $p_T$) and $|\eta^b|<2.4$\footnote{Jets are $b$-tagged with an average efficiency ($\varepsilon_{b/b}$) of $\sim 60\%$ (i.e., $\varepsilon_{b/b} = 0.85 \times \tanh(0.0025\times p_T)\times (25.0/(1+0.063\times p_T))$~\cite{CMS:2012feb}, with $p_T$ is the transverse momentum of the jet.)}. The reconstruction of $b$-jet with jets of $p_T<20$ GeV might be very challenging due to the degradation in $b$-tagging efficiency over lower transverse momentum ranges, however, it is still worthwhile to investigate how the $b$-tagging efficiency changes in response to different $p_T$ thresholds, so that we will do so below.
	
	To increase the sensitivity of our analysis, we will explore various kinematic distributions which include the invariant mass of the $b$-jets ($m_{b\overline{b}}$) as well as the constructed mass from the $\tau$ leptons decay products and the $b$-jets ($m_T^H$). These variables are typically low for signal events because the objects originate from a 125 GeV Higgs state. Conversely, they generally have higher values for background events, where the objects do not arise from a decay of a (rather light) resonance. Such kinematic features can serve to efficiently discern between signal and background events in our analysis.
	
\begin{figure}[h!]
		\centering	
		\includegraphics[scale=0.4]{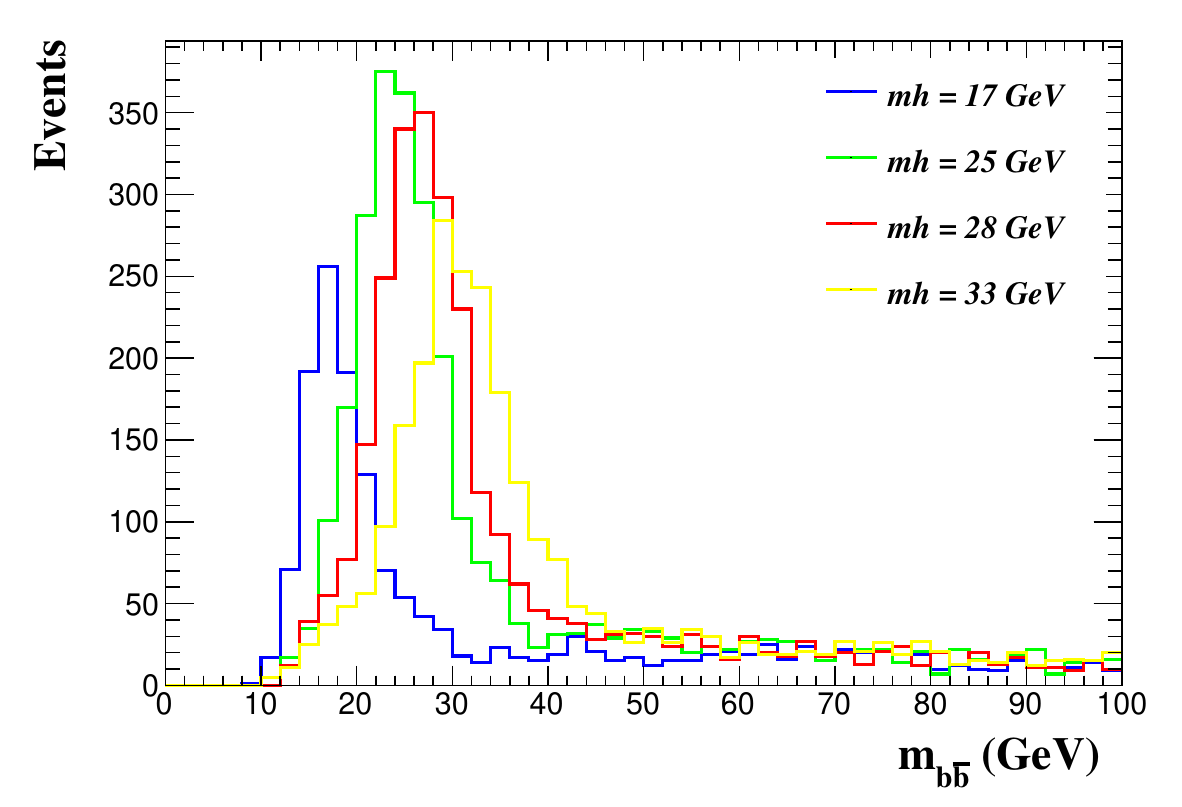}	
		\includegraphics[scale=0.4]{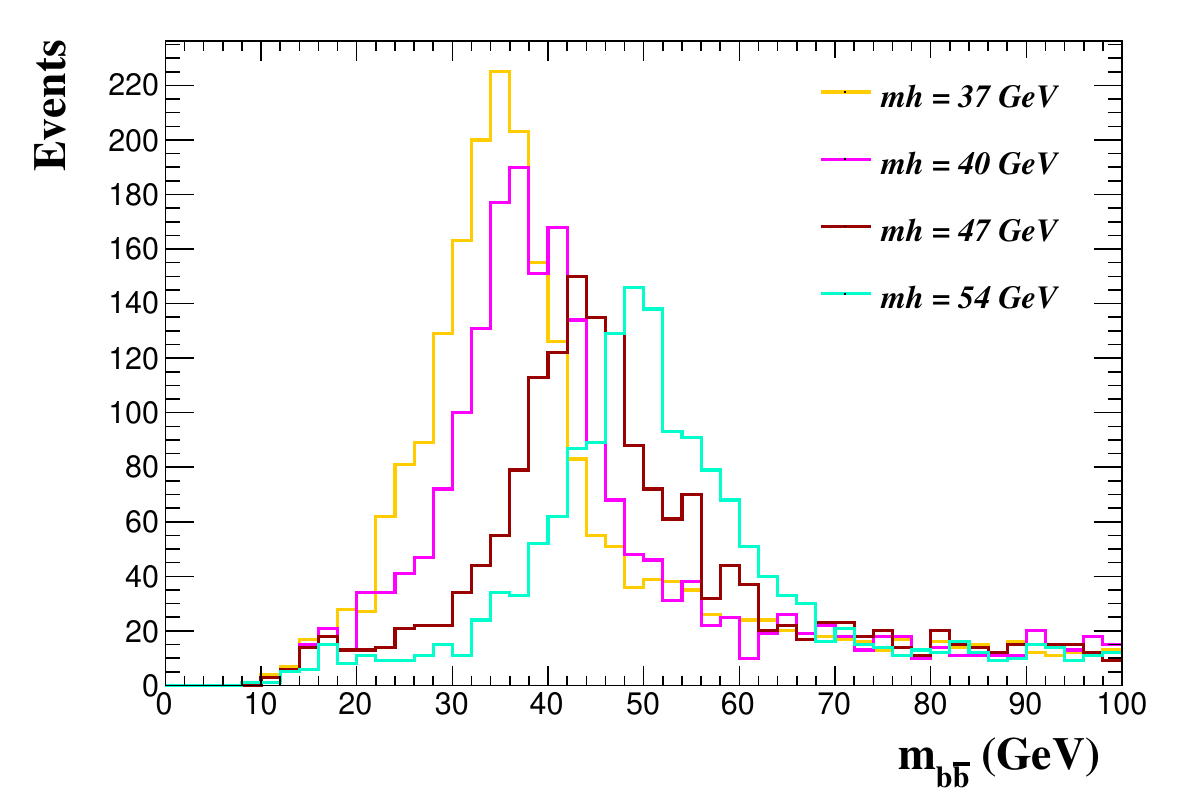}
		\caption{The distributions of $m_{b\bar{b}}$ for different BPs are shown at detector level.}	
		\label{fig8}
	\end{figure}

\begin{figure}[h!]
		\includegraphics[scale=0.4]{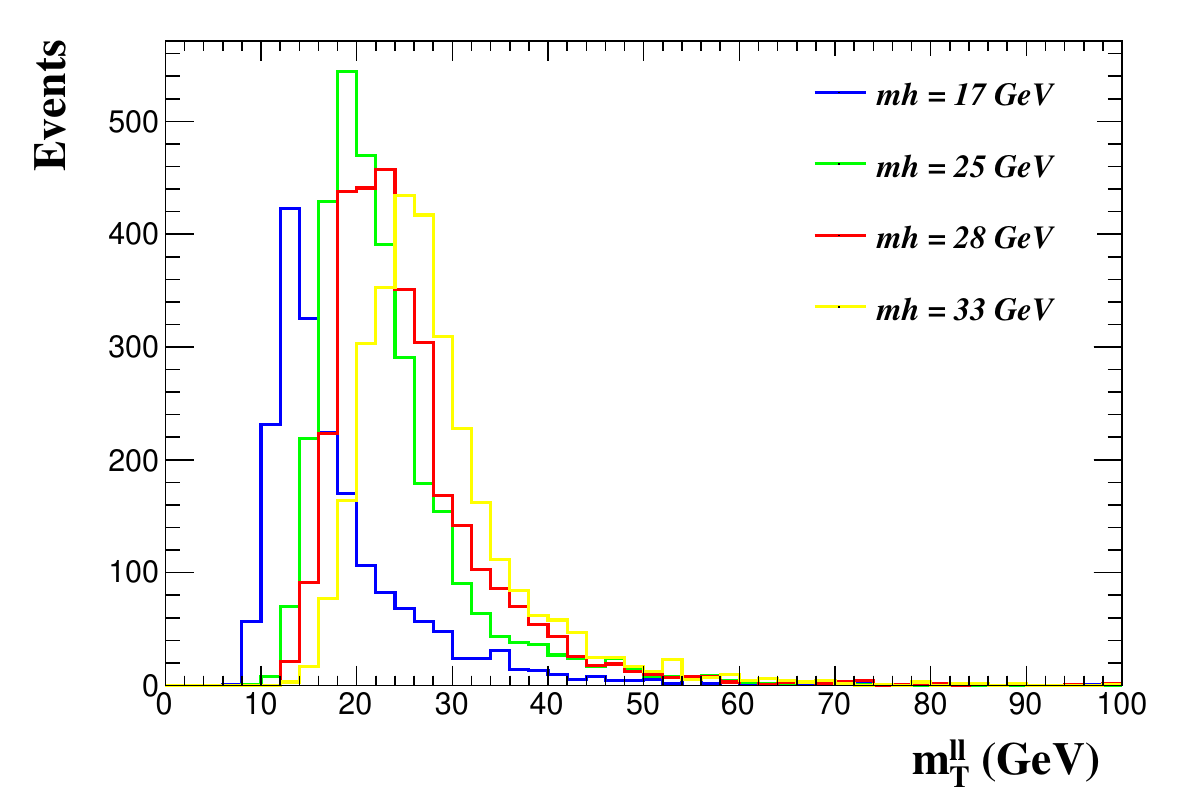}	
		\includegraphics[scale=0.4]{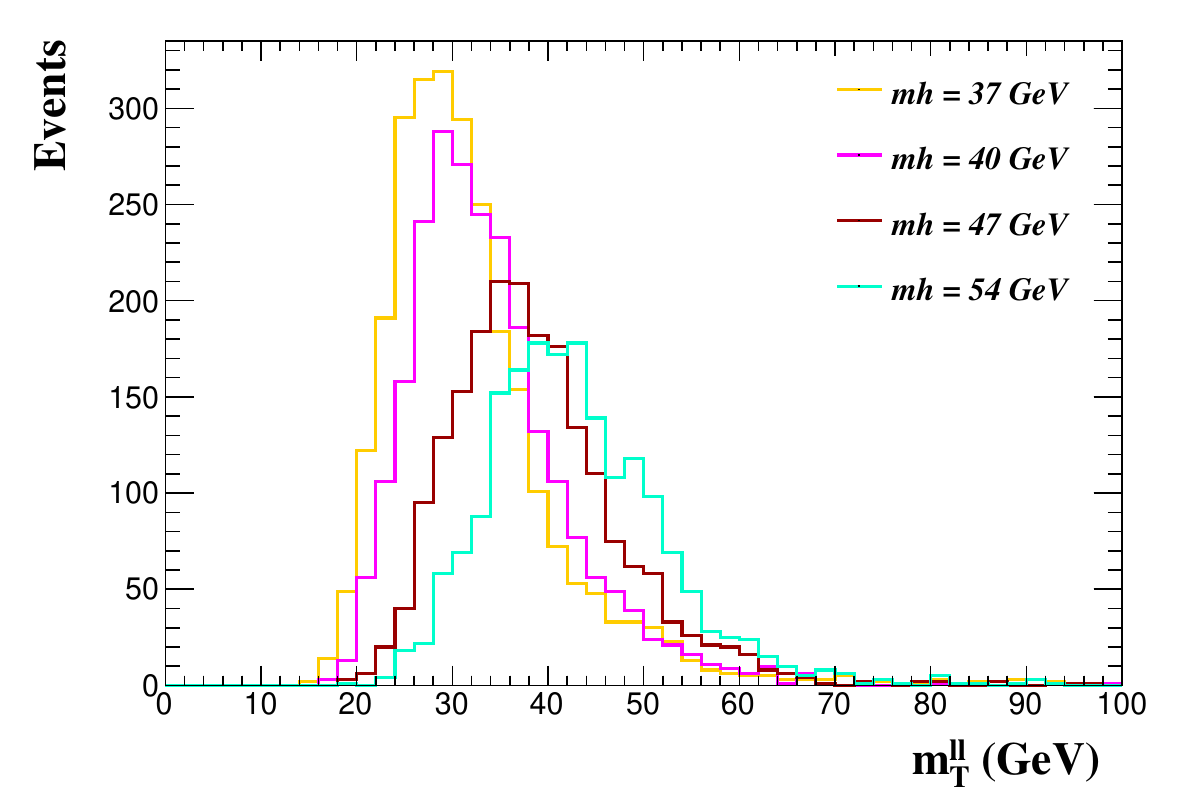}
		\caption{The distributions of $m^{ll}_T$ for different BPs are shown at detector level.}
		\label{fig9}
	\end{figure}

 Fig.~\ref{fig8} displays the invariant mass distribution of the $b$-jets for different BPs. Notably, $m_{b\overline{b}}$ aligns closely to the light Higgs mass ($m_h$) for each BP. 
	We present in Fig.~\ref{fig9} the transverse mass distributions constructed from the two charged leptons and $E_T^{\rm miss}$. As anticipated, the peak structure in these distributions is shifted from the expected light Higgs mass for the selected BPs ($m_T^{ll}\lesssim m_h$) due to the missing contribution of the neutrinos. The $m^{ l   l }_T$ variable is defined from $p_{ l  l }$ (the total four-momentum of the leptons)  and $E_T^{\rm miss}$ as
	\begin{equation}
	m^{ l  l }_T = \sqrt{ p^0_{ l  l } E^0 - |p^T_{ l  l }| |E^T| \cos(\phi_{ l  l , E_T^{\rm miss}})}.
	\end{equation}
	For the sake of convenience, we denote $E_T^{\rm miss}$ as $(E^0, E^T, p_z)$, where $p_z$ is the unknown $z$-component of the missing momentum and $E^T$ is a 2D vector defined in the $(x,y)$ plane perpendicular to the beam direction. Here, $\phi_{ l  l , E_T^{\rm miss}}$ denotes the perpendicular angle between the di-lepton system and $E_T^{\rm miss}$.  
	
\begin{figure}[h!]
		\centering	
		\includegraphics[scale=0.4]{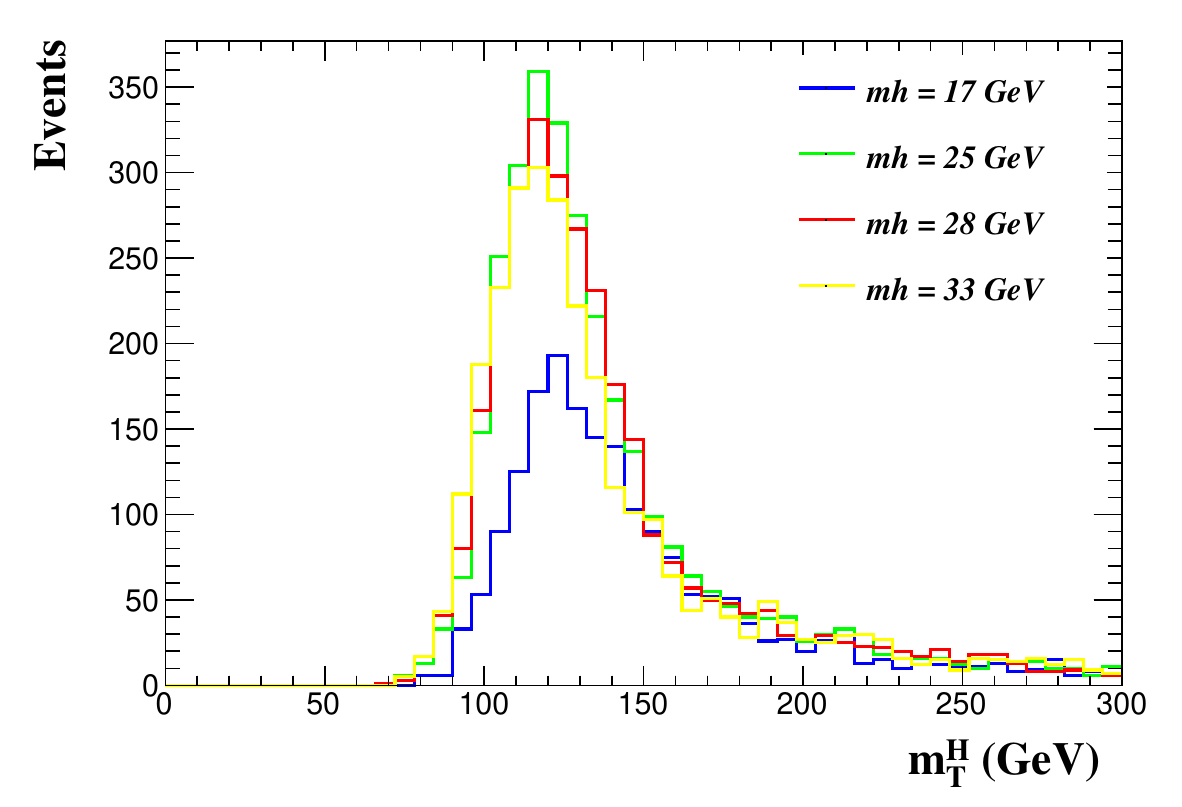}
		\includegraphics[scale=0.4]{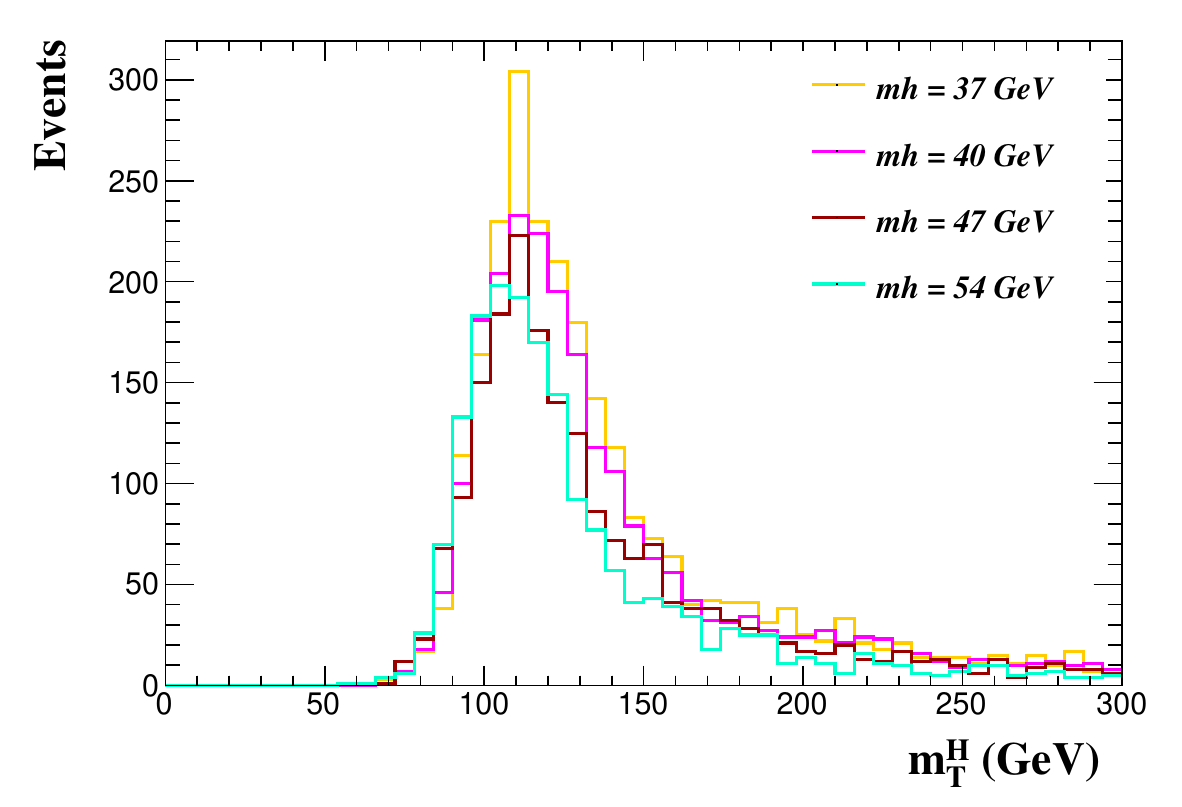}
		\caption{ The distributions of $m^{H}_T$ for different BPs are shown at detector level.}		
		\label{figa}	
	\end{figure}

	In Fig.~\ref{figa}, we show the transverse mass constructed from the two $b$-jets, the two leptons and $E_T^{\rm miss}$. The variable $m^H_T$ is defined from the two $b$-jet four-momenta $p_{b\overline{b}} = p(b) + p(\bar b)$, $p_{ l  l }$ and $E_T^{\rm miss}$.  To define $m^H_T$, we first express the visible momentum, which equals  $p_{\rm vis}= p_{b\bar b} + p_{ l   l }$, so that we have 
	\begin{equation}
	m^{H}_T = \sqrt{ p^0_{vis} E^0 - |p^T_{vis}| |E^T| \cos(\phi_{{\rm vis}, E_T^{\rm miss}})}\,,
	\end{equation}
	where  $\phi_{{\rm vis}, E_T^{\rm miss}}$ denotes the perpendicular angle between visible momentum and $E_T^{\rm miss}$. Clearly, fully reconstructing the signal can yield a significant improvement in the signal-to-background ratio. Selecting events with low $m_T^{H}$ will efficiently mitigate the background contamination arising from the $t\overline{t}$ process, which is characterised by a large missing transverse momentum. 	
	
	The requirement for  precisely two $b$-tagged jets in each event is notably a tight condition due to the soft transverse momentum ($p_T$) spectrum of $b$-jets  and the limited efficiency of the $b$-tagging algorithm. These $b$-tagging techniques~\cite{CMS:2017wtu,Cagnotta:2022hbi} operate at peak efficiency when $b$-jets possess a large transverse momentum of at least 20 GeV. However, at this threshold, a significant loss of the signal would occur particularly in scenarios involving low Higgs masses below 60 GeV  as a result of the kinematics. Therefore, to examine how the efficiencies can change, especially the $b$-tagging one, we will compare the impact of the following three cuts, which we regard as pre-selection rules:
	\begin{equation}\label{btag-cuts}
	p_T(b_1/b_2)>15/10~\text{GeV},~~~p_T(b_1/b_2)>20/15~\text{GeV},~~~p_T(b_1/b_2)>20/20~\text{GeV}.
	\end{equation}		
	The results are provided in Tabs. \ref{tab4}, \ref{tab5} and \ref{tab6}, respectively. (In these tables, to estimate the number of events, we assume that the LHC collision energy at proton-proton level is $\sqrt{s}= 13$ TeV and the integrated luminosity is that of Run 3, i.e., 300 fb$^{-1}$.) Meanwhile, the leptons from $\tau$ decays are collected with thresholds at  $p_T(e/\mu)=10/8$ GeV, which can guarantee a sufficiently large fraction of signal events to be retained.  We now discuss the effect of three pre-selections introduced in Eq.~(\ref{btag-cuts}) in turn, each of which is supplemented by the same cutflow, which has been devised following the kinematic analysis performed at detector level. \\

	In Tab.~\ref{tab4}, we adopt the following pre-selection cuts:
	\begin{equation}	p_T(b_1)>15~\text{GeV},~p_T(b_2)>10~\text{GeV},~p_T(e/ \mu)=10/8~\text{GeV}\,. \label{cuts1}
	\end{equation}
	
	As stated above, several requirements on events are enforced. The first one is that each event contains exactly two leptons $(\ell^\pm~\ell^\mp = e^\pm \mu^\mp)$. Events with lepton-pair invariant mass with 10 GeV of the $Z$-boson mass are rejected (\textgravedbl$m_Z$-veto\textacutedbl). This cut will not affect the signal but can act on background events. Following the selection of 2 tagged $b$-jets, we apply an additional selection criterion on the transverse mass $(m_T^H)$ requiring that $ 65 \,\,\textrm{GeV} < m_T^H<125 \,\,\textrm{GeV}$. This is reasonable as the Jacobian peak of the transverse mass of the SM-like Higgs boson is bound to be smaller than SM-like Higgs boson mass itself, $m_H=125 \,\, \textrm{GeV}$. 
	For the signal events, where both $b\overline{b}$ and $\tau\tau$ originate from the decay of the light Higgs, $h$, in principle, we expect the reconstructed masses $m_{b\bar b}$ and $m_{ll}^T$ to be closely correlated. In reality, a major difference between the two reconstructed masses, as shown in Fig.~\ref{fig8} and Fig.~\ref{fig9}, arises due to the $z$-component of the missing energy which is absent in $m_{ll}^T$, which prevents us from finding the precise invariant mass of the light Higgs boson decaying into $\tau\tau$. In order to quantify the difference between $m_{b\bar b}$ and $m_{ll}^T$, we have then introduced a new variable $(\Delta m_h)$ i.e., $\Delta m_h \equiv (m_{b\overline{b}}-m_{ll}^T)/m_{ll}^T$. A cut on $\Delta m_h<0.5$ would help removing background events since in the latter there is no correlation between $m_{b\bar b}$ and $m_{\tau\tau}$.
	Lastly, we set cuts on $m_T^{ l  l }$ and $m_{bb}$ to be half of $m_H$ in order to capture the SM-like Higgs boson decaying into two light Higgs bosons, $H\to hh$, in the signal events. Combining all kinematic information so far, we summarize in  Tab.~\ref{tab4} the event rates after meeting the event selection requirements, 
	
	\begin{table}[!h]
		\begin{center}
			\resizebox{0.97\textwidth}{!}{
				\begin{tabular}{|c||c|c|c|c|c|c|c|c|c|c|} 
					\hline 
					BP  &   BP1 & BP2 & BP3 & BP4 & BP5 & BP6 & BP7 & BP8 & BP9 & BP10\\
					\hline \hline
					$m_h~\text{(GeV)}$& 17.67 & 25.9 & 28.56 & 33.20 & 37.56 & 40.68 & 47.27 & 54.03 & 43.44 & 49.39 \\
					\hline
					NoE($\mathcal{L},\sigma$)& 912.86 & 727.65& 687.432& 573.3&771.74 & 769.18&1086.62& 1528.8  & 900.000 & 771.750 \\
					\hline \hline
					$e^\pm\mu^\mp$ & 156.934 & 151.874 &141.094 &114.84&146.44 & 136.2 &160.54 & 204.94 & 151.226 & 111.163 \\
					\hline 
					$m_Z$-veto &156.934 & 151.874 &141.094 &114.84&146.44 & 136.2 &160.54 & 204.94& 151.226 & 111.163 \\
					\hline
					2 $b$-jets & 33.0 & 42.88 & 39.98 & 32.32 & 38.84 & 34.94 & 40.9 & 53.2 & 38.09 &26.28 \\
					\hline
					$65~\text{GeV}< m_T^{H} < 125~\text{GeV}$ &  11.78 & 20.56 & 19.1 & 16.42 & 20.4 & 18.78 & 23.02 & 33.92 & 20.95 & 15.15\\
					\hline
					$\Delta m_{h}<0.5$ & 8.1 & 15.88 & 15.16 & 12.96 & 16.24 & 14.84 & 18.18 & 27.34 & 16.95 & 12.015\\
					\hline
					$m_{T}^{ll}<62.5$ GeV & 8.1 & 15.86 & 15.14 & 12.94 & 16.24 &14.84 & 18.18 & 27.18 & 16.63 & 11.98 \\
					\hline 
					$m_{bb}<62.5$ GeV & 8.1 & 15.86 & 15.12 & 12.94 & 16.24 & 14.76  & 18.18 & 27.04 & 16.60 &  11.97 \\
					\hline 
			\end{tabular}}
			\caption{Event rates of the signal with $\sqrt{s}=13$ TeV and integrated luminosity 300 fb$^{-1}$ for different BPs are shown as a function of our cutflow. The pre-selection cuts are as given in Eq.~(\ref{cuts1}).
			}\label{tab4}
		\end{center}
	\end{table}
	
	Tab.~\ref{tab5} presents the results based on the following pre-selection cuts, while considering the same kinematic cuts as in Tab.~\ref{tab4}:
	\begin{equation}	p_T(b_1)>20~\text{GeV},~p_T(b_2)>15~\text{GeV},~p_T(e/ \mu)=10/8~\text{GeV}.    \label{cuts2}
	\end{equation}
	Here, the same comments on the event kinematics of the signal apply as for the previous table.
	
	\begin{table}[!h]
		\begin{center}
			\resizebox{0.97\textwidth}{!}{
				\begin{tabular}{|c||c|c|c|c|c|c|c|c|c|c|} 
					\hline 
					BP  &   BP1 & BP2 & BP3 & BP4 & BP5 & BP6 & BP7 & BP8 & BP9 & BP10\\
					\hline \hline
					$m_h~\text{(GeV)}$& 17.67 & 25.9 & 28.56 & 33.20 & 37.56 & 40.68 & 47.27 & 54.03 & 43.44 & 49.39 \\
					\hline
					NoE($\mathcal{L},\sigma$)& 912.86 & 727.65& 687.432& 573.3 &771.74 & 769.18&1086.62& 1528.8  & 900.000 & 771.750 \\
					\hline \hline
					$e^\pm\mu^\mp$ & 156.934 & 151.874 &141.094 &114.84&146.44 & 136.2 &160.54 & 204.94 & 151.226 & 111.163 \\
					\hline 
					$m_Z$-veto &156.934 & 151.874 &141.094 &114.84&146.44 & 136.2 &160.54 & 204.94& 151.226 & 111.163 \\
					\hline
					2 $b$-jets & 23.9 &32.38 & 30.8 & 23.24 &29.36 &25.76 &28.56 & 40.036 & 28.125 & 18.714 \\
					\hline
					$65~\text{GeV}< m_T^{H} < 125~\text{GeV}$ & 7.76&13.72 &13.372 &11.32 & 13.82 & 12.38 & 14.58 & 22.56 & 13.890 & 9.470\\
					\hline
					$\Delta m_{h}<0.5$ & 5.232 & 11.36 & 10.77 & 8.94 & 10.76 & 9.84 & 11.92 & 17.56 & 10.973 &  7.373 \\
					\hline
					$m_{T}^{ll}<62.5$  GeV & 5.232 & 11.36& 10.76 & 8.92 &10.76 & 9.82 & 11.9 & 17.5& 10.961 & 7.363 \\
					\hline 
					$m_{bb}<62.5$ GeV &5.232 & 11.34 & 10.75 & 8.92 & 10.76 & 9.78 & 11.9 & 17.38& 10.948 & 7.363\\
					\hline 
			\end{tabular}}
			\caption{Event rates of the signal with $\sqrt{s}=13$ TeV and integrated luminosity 300 fb$^{-1}$ for different BPs are shown as a function of our cutflow. The pre-selection cuts are as given in Eq.~(\ref{cuts2}).
			} \label{tab5}
		\end{center}
	\end{table}

	In Tab.~\ref{tab6}, we show the cutflow results by adopting the pre-selection cuts: 
	\begin{equation}	p_T(b_1)>20~\text{GeV},~p_T(b_2)>20~\text{GeV},~p_T(e/ \mu)=10/8~\text{GeV}\,. \label{cuts3}
	\end{equation}	
	(Again, the same comments on the event kinematics of the signal apply as for the previous table.) 
	
	\begin{table}[!h]
		\begin{center}
			\resizebox{0.97\textwidth}{!}{
				\begin{tabular}{|c||c|c|c|c|c|c|c|c|c|c|} 
					\hline 
					BP  &   BP1 & BP2 & BP3 & BP4 & BP5 & BP6 & BP7 & BP8 & BP9 & BP10\\
					\hline \hline
					$m_h~\text{(GeV)}$& 17.67 & 25.9 & 28.56 & 33.20 & 37.56 & 40.68 & 47.27 & 54.03 & 43.44 & 49.39 \\
					\hline
					NoE($\mathcal{L},\sigma$)& 912.86 & 727.65& 687.432& 573.3&771.74& 769.18&1086.62& 1528.8  & 900.000 & 771.750 \\
					\hline \hline
					$e^\pm\mu^\mp$ & 156.934 & 151.874 &141.094 &114.84&146.44 & 136.2 &160.54 & 204.94 & 151.226 & 111.163 \\
					\hline 
					$m_Z$-veto &156.934 & 151.874 &141.094 &114.84&146.44 & 136.2 &160.54 & 204.94& 151.226 & 111.163 \\
					\hline
					2 $b$-jets & 13.6 & 20.38 & 19.02 & 15.3 & 17.86 & 16.02 & 18.34 & 23.7 & 17.39 & 11.06\\
					\hline
					$65~\text{GeV}< m_T^{H} < 125~\text{GeV}$ & 2.68 & 7.16 & 6.84 & 5.72 & 6.84 & 6.16 & 6.14 & 11.38 & 6.80& 4.19\\
					\hline
					$\Delta m_{h}<0.5$ & 1.86 & 5.5 & 5.56 & 4.5 & 5.32 & 4.8 & 5.54 & 8.2 & 5.25 & 3.13\\
					\hline
					$m_{T}^{ll}<62.5$ GeV & 1.86 & 5.5 & 5.56 & 4.48 & 5.32 & 4.8 & 5.52 & 8.2 & 5.25& 3.13\\
					\hline 
					$m_{bb}<62.5$ GeV & 1.86 & 5.5 & 5.56 & 4.48  & 5.32 & 4.78 & 5.52 & 8.12 & 5.23 & 3.13\\
					\hline 
			\end{tabular}}
			\caption{Event rates of the signal with $\sqrt{s}=13$ TeV and integrated luminosity 300 fb$^{-1}$ for different BPs are shown as a function of our cutflow. The pre-selection cuts are as given in Eq.~(\ref{cuts3}).
			}\label{tab6}
		\end{center}
	\end{table}
	
	When comparing the results given in Tabs. \ref{tab4}, \ref{tab5} and \ref{tab6}, particularly the number of events after requiring two tagged $b$-jets, it is found that the larger the $p_T$ thresholds of the b-jets, the smaller the number of signal events which can pass the pre-selection cuts. This loss is an expected outcome for a light Higgs within the sub-50 GeV range, and it aligns with our expectations. However, it is too early to conclude that with loose cuts one can maximise the sensitivity since the effects of background have to be taken into account, which is what we are going to do next.
	
	In Tab.~\ref{tab7} we show the cutflow results for the two major background processes. Here, it is noted that two mass observables, i.e.,  $m_T^H$ and $\Delta m_h$, can greatly suppress background events. Another interesting observation is that, although the number of signal events is comparatively smaller with the pre-selection cut 20/20 GeV on the two tagged $b$-jets, the background events can be almost completely removed in this case, which eventually leads to a better sensitivity to our signal process.
	\begin{table}[!h]
		\begin{center}
			\resizebox{0.87\textwidth}{!}{
				\begin{tabular}{|c||c|c|c||c|c|c||}
					\hline
					Process & \multicolumn{3}{c|}{$Zb\overline{b}$}& \multicolumn{3}{c|}{$t\overline{t}$}  \tabularnewline
					\hline 
					NoE($\mathcal{L},\sigma$) &  \multicolumn{3}{c|}{2562000}  &  \multicolumn{3}{c|}{117600} \tabularnewline
					\hline 
					$p_T(b_1/b_2)~\text{(GeV)}$ & 15/10 & 20/15 & 20/20& 15/10 & 20/15 & 20/20 \tabularnewline
					\hline
					$e^\pm\mu^\mp$ & 15836.8 & 15836.8 &  15836.8 &  61413.5  &61413.5 & 61413.5 \tabularnewline
					\hline
					$m_Z$-veto & 15801.4  & 15801.4&  15801.4& 54511.6 & 54511.6 &  54511.6\tabularnewline
					\hline
					2 $b$-jets & 1512.57 & 1059.63 & 503.558 & 16871.4 & 13778.6 & 8843.26 \tabularnewline
					\hline
					$65~\text{GeV}< m_T^{H} < 125~\text{GeV}$ & 272.439 & 154.314 & 33.2724 & 35.2954 &18.8916 & 3.087 \tabularnewline
					\hline
					$\Delta m_{h}<0.5$ GeV & 117.072 & 30.0678  & -& 17.5266 & 7.6678& -\tabularnewline
					\hline
					$m_{T}^{ll}<62.5$ GeV &  117.072 & 30.0678 & -& 14.2366 & 6.125 &-\tabularnewline
					\hline
					$m_{bb}<62.5$ GeV & 117.072 & 30.0678 & -& 14.2366 & 6.125 &-\tabularnewline
					\hline
			\end{tabular}}
			\captionof{table}{Event rates of the two dominant background processes with $\sqrt{s}=13$ TeV and integrated luminosity 300 fb$^{-1}$ as a function of our cutflow. The pre-selection cuts are as given in Eqs.~(\ref{cuts1}), (\ref{cuts2}) and (\ref{cuts3}).}
			\label{tab7}
		\end{center}
	\end{table}

	\begin{table}
	\begin{center}
		\resizebox{0.87\textwidth}{!}{
			\begin{tabular}{|c||c|c|c||c|c|c||}
				\hline
				\multirow{2}{*}{BP}& \multicolumn{3}{c|}{Significance ($\Sigma$), $\mathcal{L}=300~\text{fb}^{-1}$} & \multicolumn{3}{c|}{Significance ($\Sigma$), $\mathcal{L}=3000~\text{fb}^{-1}$} \tabularnewline 
				\cline{2-7}
				&    15/10 (GeV) & 20/15 (GeV) &  20/20 (GeV) & 15/10 (GeV) & 20/15 (GeV) &  20/20 (GeV)  \tabularnewline
				\hline
				BP1  &   0.68   &  0.81 & 1.36    &  2.15    & 2.56   &  4.30    \tabularnewline
				\hline
				BP2  &  1.30    &  1.64  & 2.34  & 4.11    &  5.18   &  7.39   \tabularnewline
				\hline
				BP3  &  1.24     &  1.57 & 2.35  & 3.92  &   4.96  & 7.43  \tabularnewline
				\hline
				BP4  &   1.07 &  1.32 & 2.11    &  3.38  &   4.17   & 6.67 \tabularnewline
				\hline
				BP5  &  1.33    &  1.57 &   2.3   & 4.20 &  4.96   &  7.27 \tabularnewline
				\hline
				BP6 &  1.22  &   1.44 & 2.18  &  3.85   & 4.55   & 6.89   \tabularnewline
				\hline
				BP7  &  1.48    &  1.71  & 2.34     &  4.68 & 5.40   & 7.39  \tabularnewline
				\hline
				BP8  & 2.14    &  2.37 & 2.84    &  6.76  & 7.49      & 8.9 \tabularnewline
				\hline 
				BP9 & 1.36 &1.59 & 2.28 &4.3 & 5.02 & 7.2 \tabularnewline
				\hline 
				BP10 & 1.0 & 1.11 & 1.76 & 3.16 & 3.51 & 5.56\tabularnewline
				\hline 
		\end{tabular}}
		\captionof{table}{Significances for our signal against the two dominant backgrounds
			with $\sqrt{s}=13$ TeV and integrated luminosity
			$300~\text{fb}^{-1}$ (left) as well as $3000~\text{fb}^{-1}$ (right).
			The pre-selection cuts are as given in Eqs.~(\ref{cuts1}), (\ref{cuts2}) and (\ref{cuts3}).}
		\label{tgg}
	\end{center}
	\end{table}
	This is explicitly shown in Tab.~\ref{tgg}, where the significances $\Sigma = \frac{\mathcal{N}_S}{\sqrt{\mathcal{N}_S+\mathcal{N}_B}}$\footnote{$\mathcal{N}_S$ and $\mathcal{N}_B$ are, respectively, the number of the signal and background events after applying the kinematic cuts discussed in the text.} for 300 fb$^{-1}$ and 3000 fb$^{-1}$ of each BP are shown. As mentioned above, the pre-selection cut 20/20 GeV can yield a better significance, so it is the one we would recommend for the actual analysis. Altogether, although (some of) these BPs might be difficult to discover or rule out at Run 3 of the LHC, all of these BPs are within full reach of the HL-LHC. 
	
	\section{Conclusions}
	The Type-I is an intriguing realisation of the 2HDM as it allows for the so-called inverted mass hierarchy scenario, wherein the Higgs boson discovered at the LHC on 4 July 2012 can be identified as the heaviest CP-even Higgs state of this construct, $H$, with a mass of 125 GeV or so and couplings to fermions and gauge bosons similar to those predicted in the SM. Such a configuration specifically implies that there is then a lighter CP-even Higgs state, $h$, into pairs of which the heavy one can decay: i.e., via $H\to hh$. Needless to say, this can be realised without contradicting any of the theoretical requirements of self-consistency of the 2HDM or current experimental results, whether coming for measurements of the discovered Higgs boson or null searches for companions to it. In fact, the latter have primarily been concentrating on other realisations of the 2HDM, where only the standard mass hierarchy is actually possible (i.e., $m_h\approx125$ GeV $<m_H$), thereby altogether missing out on the possibility of optimising searches for very light neutral Higgs states in general. Specifically, here, by looking for $H\to hh$ signals in the 2HDM Type-I,  we have concentrated on the following mass range:
	15 GeV $<m_h<m_H/2$.
	
	The production of the heavy CP-even Higgs state (the SM-like Higgs boson) at the LHC was pursued via gluon-gluon fusion, $gg\to H$, indeed, the dominant channel, while we have focused on the $ hh \to b\bar b\tau \tau$ decay pattern, where the two heavy leptons where tagged through their (different flavour) electron and muon decays.  By performing a sophisticated MC  analysis of signal versus background, we have shown that both Run 3 of the CERN machine and its HL-LHC phase can offer sensitivity to this 2HDM Type-I signal, in the presence of very low mass trigger thresholds (on the electrons and muons) already implemented for Run 3 and also possible at the HL-LHC. We have done so by adopting several BPs capturing representative $m_h$ values over the aforementioned interval after a fine scanning of the whole 2HDM Type-I parameter space, of which they are therefore representative examples amenable to further scrutiny by the LHC collaborations. Finally notice that, if the collision energy of the LHC increases from 13 TeV to 14 TeV, the production rate of the signal process $gg\to H$ can increase by $10\%$, as shown in \cite{Anastasiou:2015vya} (with the dominant backgrounds, $t\bar t$ and $Zb\bar b$, scaling similarly or less),  lending further scope to our analysis in the near future.
	
	\section*{Appendix}
\begin{figure}[h!]
		\centering
		\includegraphics[scale=0.3]{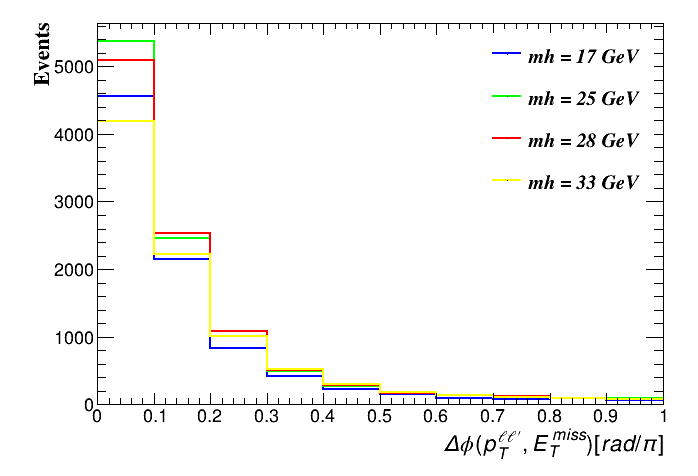}
		\includegraphics[scale=0.3]{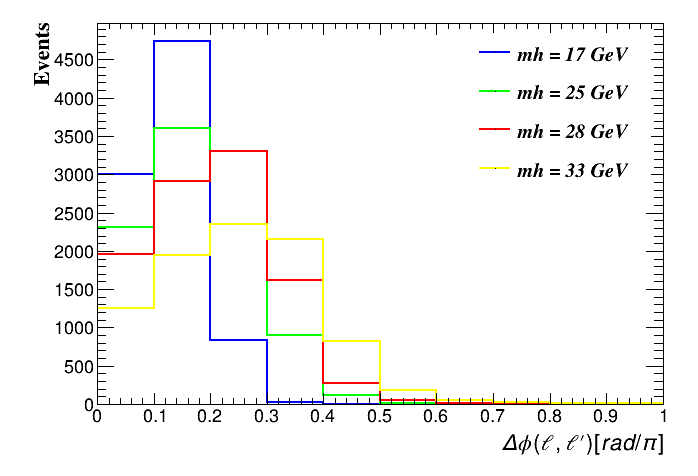}
		\caption{Opening angles distributions between the two leptons $\Delta\phi(l,l^{'})$ (left panel) and between the missing transverse momentum and the leptons $\Delta\phi(p_T^{ll^{'}}, E_T^{miss})$ (right panel) for the signal, with $(ll^{'}= e^\pm\mu^\mp)$.}
		\label{fig:enter-label}
	\end{figure}

	Additional kinematic variables can serve to further discriminate the Higgs signal from background events in the low mass range. Leptons arising from light Higgs decays demonstrate small opening angles, whereas those from $t\overline{t}$ (large missing transverse momentum) and $Z(\to\tau\tau)b\overline{b}$ (emitted back-to-back) would be large. Therefore, requiring a small opening angle between the two leptons, along with a small angle between the missing transverse momentum and the leptons would reduce the background processes. This is exemplified in Fig.~\ref{fig:enter-label}. However, we did not pursue this here.

	\section*{Acknowledgements}
	We would like to thank Sam Harper for his invaluable input and discussions around the trigger analysis. SM is supported in part through the NExT Institute and the STFC Consolidated Grant  ST/L000296/1.
	CHS-T(SS) is supported in part(full) through the NExT Institute. SS acknowledges the use of the IRIDIS High Performance Computing Facility, and associated
	support services at the University of Southampton, in the completion of this work. YW’s work is supported by the Natural Science Foundation of China Grant No. 12275143, the Inner Mongolia Science Foundation Grant No. 2020BS01013 and the Fundamental Research Funds for the Inner Mongolia Normal University Grant No. 2022JBQN080.
	QSY is supported by the Natural Science Foundation of China under the Grants No. 11875260 and No. 12275143.

	\bibliographystyle{unsrt}
	\bibliography{bibio}
	
\end{document}